\RequirePackage{snapshot}
\RequirePackage{fix-cm}
\documentclass{svjour3}                     
\smartqed  
\usepackage{graphicx}
\usepackage{array}
\usepackage{pifont}
\usepackage[export]{adjustbox}
\usepackage{natbib}

\usepackage[caption=false]{subfig}
\usepackage{amssymb}
\usepackage{booktabs}
\usepackage{tikz}
\usepackage{soul}
\usepackage{xspace}
\usepackage{tikzsymbols}
\usepackage{amsmath}

\newcommand{\figref}[1]{Figure~\ref{#1}\xspace}
\newcommand{\secref}[1]{Section~\ref{#1}\xspace}
\newcommand{\tabref}[1]{Table~\ref{#1}\xspace}

\definecolor{darkred}{rgb}{0.75,0,0}

\newcommand{\cmark}{\ding{51}}%
\newcommand{\xmark}{\ding{55}}%
\newcommand{\roundbox}[1]{
	\begin{center}
		\begin{tikzpicture}
		\node[draw=black, rectangle, rounded corners](box){
			\begin{minipage}{0.9\columnwidth}
			#1
			\end{minipage}
		};
		\end{tikzpicture}
	\end{center}
}

\begin{document}
	
	\title{On the Time-Based Conclusion Stability of Cross-Project Defect Prediction Models}
	
	
	
	\author{Abdul Ali Bangash \and
		Hareem Sahar \and
		Abram Hindle \and
		Karim Ali \and
	}
	
	
	
	\institute{Abdul Ali Bangash \at
		Department of Computing Science\\
		University of Alberta, Edmonton, AB, Canada \\
		\email{bangash@ualberta.ca}           
		\and
		Hareem Sahar \at
		Department of Computing Science\\
		University of Alberta, Edmonton, AB, Canada \\
		\email{hareeme@ualberta.ca}           
		\and
		Abram Hindle \at
		Department of Computing Science\\
		University of Alberta, Edmonton, AB, Canada \\
		\email{abram.hindle@ualberta.ca}           
		\and
		Karim Ali \at
		Department of Computing Science\\
		University of Alberta, Edmonton, AB, Canada \\
		\email{karim.ali@ualberta.ca}           
	}

	\date{Received: date / Accepted: date}

	\maketitle

	\begin{abstract}
		Researchers in empirical software engineering often make claims based on observable data such as defect reports. Unfortunately, in many cases, these claims are generalized beyond the data sets that have been evaluated. Will the researcher's conclusions hold a year from now for the same software projects? Perhaps not. 
		Recent studies show that in the area of Software Analytics, conclusions over different data sets are usually inconsistent. In this article, we empirically investigate whether \emph{conclusions} in the area of cross-project defect prediction truly exhibit \emph{stability} throughout time or not.
		Our investigation applies a time-aware evaluation approach where models are trained only on the past, and evaluations are executed only on the future. Through this time-aware evaluation, we show that depending on which time period we evaluate defect predictors, their performance, in terms of F-Score, the area under the curve (AUC), and Mathews Correlation Coefficient (MCC), varies and their results are not consistent. The next release of a product, which is significantly different from its prior release, may drastically change defect prediction performance. Therefore, without knowing about the conclusion stability, empirical software engineering researchers should limit their claims of performance within the contexts of evaluation, because broad claims about defect prediction performance might be contradicted by the next upcoming release of a product under analysis.                                                                                                 
	\end{abstract}
	
	\section{Introduction}
	\label{sec:intro}
	
	\emph{Defect prediction models} are trained for predicting future software bugs using historical defect data available in software archives and relating it to predictors such as structural metrics \citep{CK1994, Martin1994, tang1999}, change entropy metrics \citep{hassan2009}, or process metrics \citep{mockus2000metrics}. The accuracy of defect prediction models is estimated using defect data from a specific time period in the evolution of software, but the models do not necessarily generalize across other time periods.
	
	\emph{Conclusion stability} is the property that a conclusion, i.e., the estimate of performance, remains stable as contexts, such as time of evaluation, change. For example, if the conclusion of a current evaluation of a model on a software product is the same as that of an evaluation done a year ago, then we consider that conclusion to be stable. A lack of conclusion stability would be if the model's performance is inconsistent with itself across time. Instead of over generalizing our conclusions beyond the period of evaluation, if we claimed the model's performance was within the period of evaluation, our claim would still hold. 
	
	Prior work \citep{lessmann2008bench,menzies2010,turhan2012} examined various factors affecting the conclusion stability of defect prediction models. However, none explored the conclusion stability across time. The goal of this paper is to investigate conclusion stability of cross-project defect prediction models (trained and tested using data from different projects) and understand how their performance estimates, measured using F-Score, Area under the Curve (AUC), Matthews Correlation Coefficient (MCC), and G-measure vary across different time periods. 
	In our evaluation, we carefully consider the time-ordering of versions and ensure our models do not involve \emph{time-travel}. \emph{Time-travel} is a colloquial term to describe models that should be time sensitive but are trained on future knowledge that should not be known for predicting defects in the past. 
	
	Existing defect prediction studies fail to avoid \emph{time-travel} because of the choice of a cross-validation evaluation methodology which,
	
	1) Randomly splits data into partitions and uses these partitions for training and testing, irrespective of the chronological order of data. 
	
	2) Reports the mean performance metrics without specifying the evaluated time period, and assumes the performance generalizes over all time periods.
	
	The main drawback of this methodology is that the defect prediction models often get trained on future data which is not available, in reality, at the time of training. For example, due to cross-validation, a version released in 2010 may be used for training a model that predicts defects for a version released in 2009. This situation is explained in \tabref{table:tsExample} that shows a cross-validation evaluation for three software releases ($i,j,k$), each from three different projects, released between 2008 and 2010. The table shows that not all Training (Tr) and Test combinations are realistic for building defect prediction models, as some will lead to models that are time insensitive (trained on future data). For instance, a case where Tr set = \{j\} and Test set = \{i\}, the evaluation seemingly have engaged in \emph{time-travel}.

	\citet{rakha2018} refer to such evaluation as \textit{classical evaluation}, whereas \citet{hindle2019query} call it \textit{time-agnostic}. 
	Many claim that ignoring time provides highly unrealistic performance estimates \citep{tan2015online, rakha2018, hindle2019query}, yet, there are several just-in-time based approaches that only consider release order for within project defect prediction \citep{huang2017supervised, EMSE6effortaware}, but engage in \emph{time-travel} in cross project defect prediction settings \citep{EMSE6effortaware, kamei2016studying, yang2015deep}.\footnote{\citet{yang2015deep} used 10-fold cross-validation in their study. On the other hand, \citet{EMSE6effortaware} used time-wise cross-validation for within-project models, however, in cross-project prediction they trained on one project and tested on another project without ordering the data set time-wise. \citet{kamei2016studying} trained JIT cross-project models using the data from one project and tested the prediction performance using the data from every other project, irrespective of their time order.} 

	
	\begin{table}[t]
		\caption{An example illustrating three cross-validation settings (a=1/1, b=2/1, c=1/2) of 
			three releases of different projects over a period of three years (i-2008, j-2009, k-2010). }
		
		\begin{center}
			
			\subfloat[][Cross-validation (1/1) having 1 release in Training set and 1 release in Test set]{
				\begin{tabular}{ccc}
					\toprule
					\multicolumn{2}{c}{\bf Cross Validation} & \textbf{Training/Test}\\	
					
					\cmidrule(lr){1-2} \cmidrule(lr){3-3}
					Training set & Test set & Time-travel \\ \midrule
					\{i-2008\} &    \{j-2009\} & \xmark \\ 	
					\{i-2008\} &	\{k-2010\} & \xmark \\	
					\{j-2009\} &	\{i-2008\} & \cmark \\	
					\{j-2009\} &	\{k-2010\} & \xmark \\	
					\{k-2010\} &	\{i-2008\} & \cmark \\	
					\{k-2010\} &	\{j-2009\} & \cmark \\	\bottomrule
					
				\end{tabular}
			}
			
			\subfloat[][Cross-validation (2/1) having 2 releases in Training set and 1 release in Test set]{
				
				\begin{tabular}{ccc}
					\toprule
					\multicolumn{2}{c}{\bf Cross Validation} & \textbf{Training/Test}\\	
					
					\cmidrule(lr){1-2} \cmidrule(lr){3-3}
					Tr set & Test set & Time-travel \\ \hline
					\{i-2008, j-2009\} & \{k-2010\} & \xmark \\ 	
					\{j-2009, k-2010\} &	\{i-2008\} & \cmark \\	
					\{k-2010, i-2008\} &	\{j-2009\} & \cmark \\	\bottomrule
					
				\end{tabular}
				
			}

			\subfloat[][Cross-validation (1/2) having 1 release in Training set and 2 releases in Test set]{
				\begin{tabular}{ccc}
					\toprule
					\multicolumn{2}{c}{\bf Cross Validation} & \textbf{Training/Test}\\	
					
					\cmidrule(lr){1-2} \cmidrule(lr){3-3}
					Tr set & Test set & Time-travel \\ \hline
					\{i-2008\} & \{j-2009, k-2010\} & \xmark \\ 	
					\{j-2009\} & \{i-2008, k-2010\} & \cmark \\	
					\{k-2010\} & \{i-2008, j-2009\} & \cmark \\	\bottomrule
					
				\end{tabular}
			}
		\end{center}
		
		\label{table:tsExample}
	\end{table}

	In this paper, we evaluate five cross-project defect prediction approaches using the publicly available Jureczko dataset \citep{Jureckzo}, and show that data from different time periods leads to varying conclusions. In our evaluation, we strictly consider the chronological order of data and propose four generic time-aware configurations that can be used to split the data set into training and testing. 
	The purpose of proposing these configurations is to make the experiment performance-wise scalable for evaluating other approaches in which running all possible Tr-Test set combinations is expensive, such as duplicate bug reports retrieval involving extensive string matching \citep{hindle2019query}.
	
	Our results indicate that the evaluated cross-project defect prediction approaches do not have perfect stability in their conclusions and \emph{time-travel} produces false estimates of performance. Therefore, while conducting defect prediction studies, researchers should not engage in \emph{time-travel} and also avoid over generalizing their conclusions, but instead couch the claims of performance within the contexts of evaluation. 
	To summarize, the main contributions of this paper are:
	\begin{itemize}
		\item A methodology for time-aware evaluation of defect prediction approaches; 
		\item A case study of conclusion stability in cross-project defect prediction with respect to time;
		\item A comparison of the performance rankings of five cross-project defect prediction approaches using time-aware evaluation with performance of time agnostic evaluation;
		\item Guidelines for researchers and practitioners for the time-aware evaluation of defect prediction models.
	\end{itemize}
	\section{Related Work}
	
	Software defect prediction has a plethora of approaches with the earliest proposals dating back to the 1990s where linear regression models based on Chidamber and Kemerer (CK) metrics \citep{CK1994} were used to determine the fault proneness of classes \citep{basili1996}. A number of metrics have been used since then as indicators of software quality such as previous defects \citep{zimmermann2007}, process metrics \citep{hassan2009, prem2013}, and churn metrics \citep{nagappan2005}. Within project defect prediction (WPDP) uses data from the same project for training and testing whereas in cross project defect prediction (CPDP), training and testing data comes from different projects. 
	Several approaches for both WPDP \citep{turhan2009relative, basili1996} and CPDP \citet{zimmermann2009cross, menzies2013better, nam2013transfer} are available in the literature. There have also been benchmark studies on both types of defect prediction \citep{Ambros2012, Herbold}. WPDP approaches have better performance while CPDP approaches are likely transferable to other projects with certain limitations \citep{zhang2014}. \citet{herbold2017mapping} conducted a systematic mapping of defect prediction literature with a focus on cross project defect prediction approaches. They identified that the results of studies are not comparable due to the lack of use of common data sets and experimental setups. 
	
	In their follow up work, \citet{Herbold} replicated 24~defect prediction approaches using 5~publicly available data sets and multiple learners. Their goal was to benchmark the defect prediction approaches using common data sets and metrics so that state-of-the-art approaches can be ranked according to their performance using Area under Curve (AUC), F-Score, G-measure, and Matthews Correlation Coefficient (MCC) metrics. Jureczko \citep{Jureckzo} is one of the well-known defect prediction data sets which was also used in the benchmarking study. It originally contains open-source, proprietary and academic projects but \citet{Herbold} used only 62 versions of several open-source and academic projects. 
	
	Prior to this paper, conclusion stability has been analyzed by several researchers. \citet{lessmann2008bench} and \citet{menzies2010} investigated the effect of classifiers, trained using same data, on the quality of prediction models whereas \citet{ekanayake2012,ekanayake2009} investigated the effect of data set and concept drift respectively. \citet{lessmann2008bench} found statistically significant difference among the performance of two classifiers and \citet{menzies2011local} observed inconsistent conclusions for different clusters within the same data. Inspired by this prior work, another set of experiments were conducted by \citet{Ambros2012} to rank approaches across several data sets following a statistically sound methodology. \citet{mcintosh2017fix} investigated the time-based conclusions of just-in-time defect prediction models and found that their descriminatory power can change over time. In more recent work, \citet{Chakkrit2018} concluded that parameter optimization can significantly impact the performance stability, and ranking of defect prediction models. This view is similar to Menzies' view who argued that a learner tuned to a particular evaluation criterion, performs best for that criterion, hence it shall be critically chosen \citep{menzies2010}. 
	
	\citet{chakkrit2015mislabel} in their work show that issue report mislabelling significantly impacts the defect prediction models. In a later comparison study, \citet{chakkrit2017modelValidation} concluded that the choice of model validation technique for defect prediction models can also affect performance results. \citet{tan2015online} identified that cross-validation produces false precision results for change classifications and addressed the problem using time-sensitive and online change classifications. Their emphasis is on removing imbalances in data using re-sampling techniques for better change classifications. \citet{turhan2012} also studied the conclusion instability caused due to data set shift but their focus was not specific to defect prediction, rather on software engineering prediction models in general. Similarly \citet{EMSE5bellwethers} show that there can be large differences in conclusions depending on different source data sets and suggest mitigating the problem with the help of bellwethers. Bellwethers seem to restrain instability but based on the results of our study, we consider it of utmost importance to keep regard of time  while finding out the bellwether project. However, we believe this work complements our work. 
	
	Time-agnostic evaluation has been criticized as unrealistic by \citet{hindle2019query} who argue that the results based on a time-agnostic evaluation might not be applicable to any real-world context. \citet{EMSE6effortaware} hold a similar view and motivated by \citet{ZimmermanSlidingwindow} they adopted a time-wise cross-validation within projects for evaluating the prediction effectiveness of unsupervised models. However, in their cross-project defect prediction setting, they seem to be  time travelling again. Instead of using their approach we propose four time-aware configurations to avoid discarding some of the valid models that time-wise cross-validation will not generate. \citet{jimenez2019} assessed the impact of disregarding temporal constraints on the performance of vulnerability prediction models and found that the otherwise highly effective and deployable results quickly degrade to an unacceptable level when realistic information is considered. Their work is limited to the prediction of vulnerabilities though, which are just a subset of defects. \citet{rakha2018} also claim that time-agnostic evaluation overestimates performance. They argue that the range of performance estimates, rather than a single value should be reported.
	
	\section{Methodology}
	
	In this section, we explain a time-aware evaluation methodology that we follow for building the cross-project defect prediction models that do not engage in time travel. To avoid time-agnostic evaluation in future, researchers can employ this proposed methodology for the evaluation of their defect prediction techniques.
	
	\subsection{Select techniques to evaluate}
	The first step is to select techniques for validation, and these can either be newly proposed techniques or existing defect prediction proposals. In general, defect prediction techniques can be selected from a broad category of \emph{within project defect prediction techniques} (WPDP) or \emph{cross project defect prediction techniques} (CPDP). As the name suggests, WPDP uses the same project in training and testing, whereas CPDP is across different projects. CPDP has several variants including strict CPDP, mixed CPDP, and pair-wise CPDP \citep{herbold2017mapping}. In strict CPDP, there is a strict distinction between the projects used in training and testing. This restriction implies that none of the projects used for training the model remain part of the testing data so that information from same context does not mix up. Contrarily, in mixed CPDP, some releases of a project are used for training while others are used for testing. In pair-wise CPDP, a separate model is trained using each project release, and their performance is averaged for estimating the actual performance. 
	
	\subsection{Extract software defect prediction metrics with dated releases}
	Existing software systems with issue trackers can be used to extract software defect prediction metrics and post-release defects via mining software repositories. Extraction methodologies discussed in prior work \citep{extraction2005, extraction2003, extraction2007} can be leveraged for the purpose of gathering data. We can alternatively benefit from existing defect data sets used by prior studies for evaluating the technique. One has to make sure that the data set contains releases that have dates or time-stamps. Alternatively, if versions are specified, one can extract and use version release dates. For example, if the data set contains commit history ids, bug report ids, and version release tags, we can extract version release dates from these factors. Before moving on to the next step, one has to label the defect data set instances with dates or timestamps. 

	\begin{figure*}
		\centering
		\includegraphics[width=\textwidth]{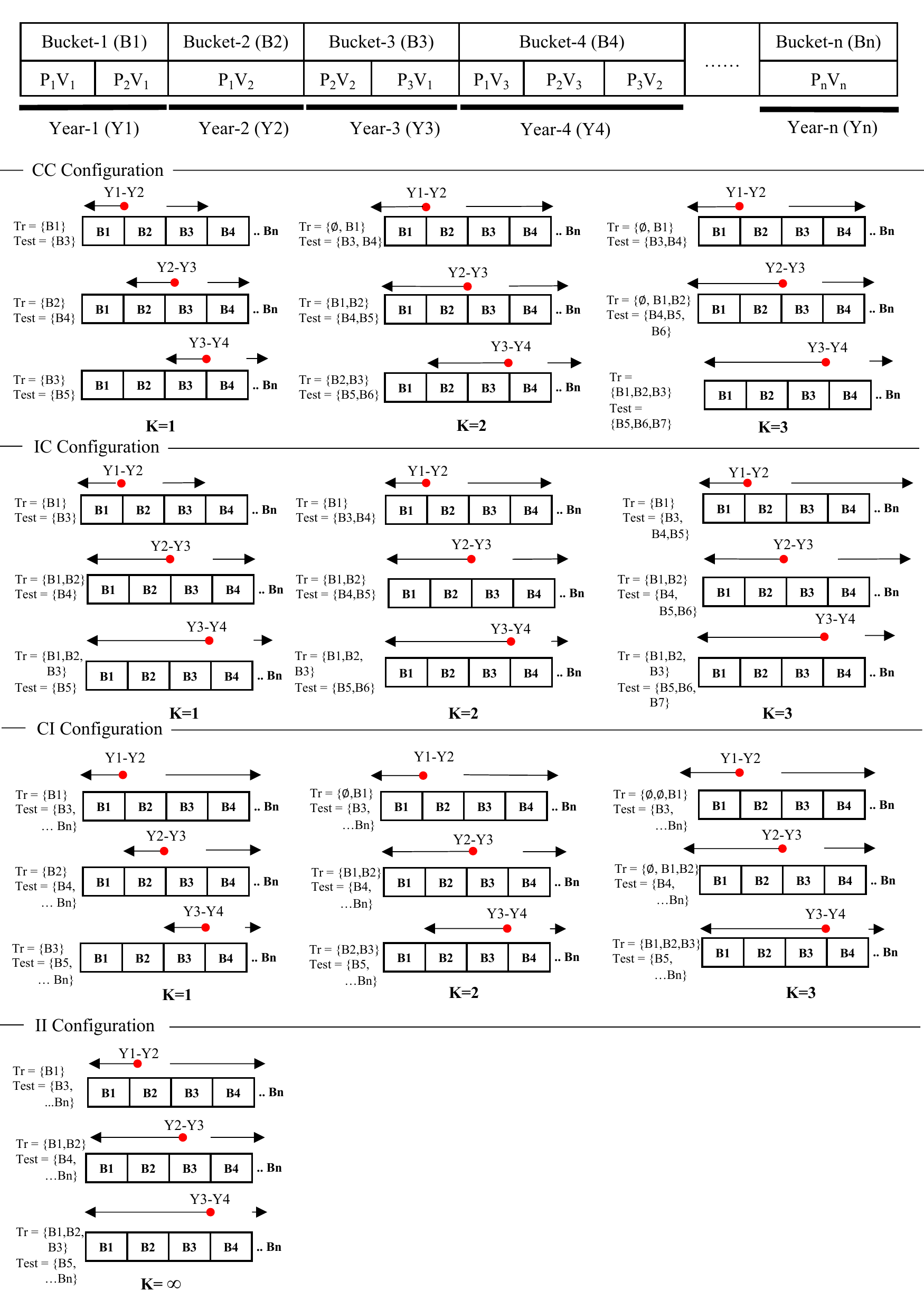}
		\caption{Generating Training (Tr) and Test (Test) pairs using four time-aware configurations: Constant-Constant (CC), Increasing-Constant (IC), Constant-Increasing (CI), Increasing-Increasing (II). Pn refers to Project number, Vn refers to Version number,  Yn refers to (Year number), and K is Window size and decides the number of time buckets that are used in training and testing. $\phi$ in II means that Window size does not matter in that configuration.}
		\label{fig:MainExample}
	\end{figure*}

	\subsection{Sort and Split project versions into time buckets}
	In this step, the defect data set is first sorted according to the time available in the form of version dates,
	and then split using N split points. A \textbf{split point} is the \emph{reference point in time} that partitions the defect data into time-buckets, and it is chosen such that the data is partitioned into a day, month, or year granularity. Consequently, each time-bucket spans days, months, or years of releases. 
	
	
	\figref{fig:MainExample} illustrates how an example N-year long data set is divided into N buckets using split point at one year granularity. Bucket-1 is formed starting from the oldest project version until the first split, so it contains project versions spanning a year. Bucket-2 contains one year data between first and second split, and so on. In this way, all versions of all projects released within one specific year fall into the bucket representing that year. The choice of window size, and hence the bucket granularity, may vary depending on the available data and time information. 
	If there were no project versions in, for example, Year-2 in \figref{fig:MainExample}, then Bucket-2(B2) would also remain empty. On the other hand, in the current example, one version ($V_{2}$) of a project ($P_{1}$) was released in Year-2, and, therefore, it is included in B2. Therefore,
	one may observe that the number of projects and versions in each bucket are unequal. 
	
	These split points allow the software versions before a certain split to be used for training set while any versions after that split form the test set. 
	Unlike cross-validation there is no time-travelling in such evaluation because the buckets are ordered by time. 
	Notice that a lower granularity spreads the data set well across the timeline and a great number of data points are available for constructing and evaluating the defect prediction models. For the rest of this paper, we will refer to these time ordered buckets as a \emph{time-series} data set.

	\begin{table}
		
		\caption{An example illustrating four time-aware settings (a=CC, b=IC, c=CI, d=II) of three releases of three different projects over a period of three years (i-2008, j-2009, k-2010). $\phi$ = empty set representing no release available at that time. $\infty$ = max window size possible.}
		
		\begin{center}
			\subfloat[][Configuration Constant-Increasing (CC)]{
				\begin{tabular}{cccc}
					\toprule
					\textbf{Tr set} & \textbf{Test set} & \textbf{Split point} & \textbf{Window size} \\ \midrule
					\{i-2008\} & \{j-2009\} & 2008-2009 & 1 \\ 
					\{j-2009\} & \{k-2010\} & 2009-2010 & 1 \\ 
					\{$\phi$, i-2008\} & \{j-2009, k-2010\} & 2008-2009 & 2 \\ 
					\{i-2008, j-2009\} &  \{k-2010, $\phi$\} & 2009-2010 & 2 \\ \bottomrule
					
				\end{tabular}
			}
			
			\subfloat[][ Configuration Increasing-Constant (IC)]{
				\begin{tabular}{cccc}
					\toprule
					\textbf{Tr set} & \textbf{Test set} & \textbf{Split point} & \textbf{Window size} \\ \midrule
					\{i-2008\} & \{j-2009\} & 2008-2009 & 1 \\ 
					\{i-2008, j-2009\} & \{k-2010\} & 2009-2010 & 1 \\
					\{i-2008\} & \{j-2009, k-2010\} & 2008-2009 & 2\\ 
					\{i-2008, j-2009\} & \{k-2010, $\phi$\} & 2009-2010 & 2 \\ \bottomrule
					
				\end{tabular}
			}
			
			\subfloat[][Configuration Constant-Increasing (CI)]{
				\begin{tabular}{cccc}
					\toprule
					\textbf{Tr set} & \textbf{Test set} & \textbf{Split point} & \textbf{Window size} \\ \midrule
					\{i-2008\} & \{j-2009, k-2010\} & 2008-2009 & 1 \\ 
					\{j-2009\} & \{k-2010\} & 2009-2010 & 1 \\ 
					\{$\phi$, i-2008\} & \{j-2009, k-2010\} & 2008-2009 & 2 \\ 
					\{i-2008, j-2009\} & \{k-2010, $\phi$\} & 2009-2010 & 2 \\ \bottomrule
					
				\end{tabular}
			}
			
			\subfloat[][Configuration Constant-Increasing (II)]{
				\begin{tabular}{cccc}
					\toprule
					\textbf{Tr set} & \textbf{Test set} & \textbf{Split point} & \textbf{Window size} \\ \midrule
					\{i-2008\} & \{j-2009, k-2010\} & 2008-2009 & $\infty$ \\ 
					\{i-2008, j-2009\} & \{k-2010\} & 2009-2010 & $\infty$ \\ \bottomrule
					
				\end{tabular}
			}
		\end{center}	
		
		\label{table:tsConfigs}
		\vspace{-2em}
	\end{table}
	
	\subsection{Generate Training-Test pairs from time buckets}
	In this step, we use the time-series data set to generate multiple \emph{Training-Test} (Tr-Test) pairs following four time-aware configurations. 
	\figref{fig:MainExample} provides a high-level overview of these configurations where the time granularity of buckets is one year, and each bucket contains multiple project versions. 
	In each configuration, the split point divides the data into two parts: past and future. The red dot represents a split point in \figref{fig:MainExample}. 
	The buckets containing project versions before the split point form the past of a data set and will be considered for training (Tr) while those after the split point (after skipping one bucket) form the future and are used for testing (Test). The reason for skipping one bucket is to reduce the possible chances of time-travel within the instances of training and test data and to allow some time for buggy changes in the
	training set to be discovered and fixed. This gap can vary and should ideally be equal to the time that it takes for a bug to be reported and fixed.
	We further employ \textbf{window size} to select the \textit{number of time-buckets} to be used for generating Tr-Test pairs. 
	The window size also has a granularity in terms of the number of time buckets, e.g., a window size of one corresponds to one year of data in our example.
	Consequently, the Tr-Test set size, i.e., the number of project versions in training and test set, varies as window size changes: number of project versions is not constant in every bucket.
	
	To explain the four configurations, we use the example of \tabref{table:tsExample} introduced earlier in \secref{sec:intro} and present Tr-Test pairs corresponding to the four time-aware configurations in \tabref{table:tsConfigs}. $\phi$ in the table represents an empty set for the cases when window size exceeds the number of buckets available in the data set for Tr or Test set. Unlike \figref{fig:MainExample}, for the sake of brevity, the gap between Tr-Test pairs in \tabref{table:tsConfigs} is not shown.
	
	\textit{\textbf{Configuration 1 | Constant-Constant (CC):}}  In this configuration, the Tr and Test set are populated according to the window size. 
	At each split point with a constant window size $K$, we take $K$ time-buckets before the split point for Tr set and an equal number of buckets after one bucket gap of the split point for Test set, as shown in \figref{fig:MainExample}. This Tr set and Test set forms a Tr-Test pair. The window size is increased once the Tr-Test pairs over all split points are generated. As a result, we get one Tr-Test pair corresponding to each value of window size and split point.
	
	The process of generating Tr-Test pairs is repeated until all possible pairs corresponding to each split point and window size are generated.
	There can be cases where an equal number of buckets before and after the split point are not available, for example, if we consider CC configuration's K=3 at split Y1-Y2 in \figref{fig:MainExample} there is only one bucket available for training. To ensure consistency in generating configurations, we consider as many buckets as available at such split points, hence our Tr set~=~\{$\phi$,B1\}. This configuration is similar to the evaluation of \citet{rakha2018} except that they employed tuning.

	\textit{\textbf{Configuration 2 | Increasing-Constant (IC):}} At each split point in this configuration, the Test set is populated with $K$ time buckets after skipping one bucket after the split point, where $K$ is the window size. While the Tr set is populated with all the time-buckets available before the split point. Same as CC, the window size is increased once the Tr-Test pairs over all split-points are generated. Considering each split point and current window size value referred to as $K$ in \figref{fig:MainExample}; we take all time-buckets before the split point for Tr and $K$ number of buckets after skipping one bucket after the split point for Test. The example Tr-Test pairs corresponding to each value of window size and split point are shown in \figref{fig:MainExample}. 
	
	\textit{\textbf{Configuration 3 | Constant-Increasing (CI):}} Contrary to IC, at each split point in this configuration, the Tr set instead of Test set is populated with $K$ time buckets before the split point, where $K$ is the window size. Whereas same as CC and IC, the window size is increased once the Tr-Test pairs over all split-points are generated. Considering each split point and current window size value referred $K$; we take $K$ number of buckets before the split point for Tr while all time-buckets after skipping one bucket after the split point for Test. The example Tr-Test pairs corresponding to each value of window size and split point are shown in \figref{fig:MainExample}.

	\textit{\textbf{Configuration 4 | Increasing-Increasing (II):}}	In II, the window size does not matter because at each split point, the Tr-Test pairs are generated by taking all the buckets before split point for training and all those after skipping one bucket after the split point for testing. We set window size or $K$ in this configuration to infinity as that is theoretically the maximum possible window size.
	
	Each configuration serves a different purpose, and depending on the context one configuration is a more  appropriate choice than the other. For example, the quality assurance team wanting to test the next due release of a project against the entire past may use IC or II configurations. The CI configuration is more useful in cases where a major release in the past has entirely changed the system, and the developers want to test their system since then. CC and II configurations might benefit researchers who are trying to evaluate the defect prediction methodologies, so they can evaluate and compare the performance of defect prediction approaches. It is still a matter of research to find out which configuration is a better choice for what kind of environment. However, we employ all four configurations in our experiments.

	\subsection{Build prediction models and evaluate performance} 
	Each technique applies certain treatment on the instances in the training and test set before building the model. For example, one technique may apply log transformation on the training set, while another may use K-Nearest Neighbours (KNN) relevancy filtering. Therefore, we apply the treatment proposed by a defect prediction technique to all the Tr and/or Test sets generated in the previous step and then build a prediction model from each Tr set. We then evaluate that model on each project version in the Test set and calculate the mean performance. 
	For example, in \figref{fig:MainExample}, consider \textbf{IC} configuration's second setting, where K=1 at split point Y2--Y3, the training set Tr is {B1,B2} and Test set is {B4}. Given that B4 consists of three projects {P1V3, P2V3, P3V2}. For this setting, we will train one prediction model and evaluate it on three separate test sets: trained on Tr={B1,B2} and tested on Test={P1V3}, then on Test={P2V3}, and finally on Test={P3V2}. Similarly, if there are multiple versions of a same project in the test bucket, we evaluate each version separately. If there was a P3V4 in B4, then we would test that separately as well.

	\section{Experimental Setup}
	In this section, we employ the proposed time-aware configuration settings to investigate the conclusion stability of cross-project defect prediction approaches.
	
	\subsection{Select techniques to evaluate} In this work, we do not propose a new defect prediction approach. Instead, we re-evaluate existing cross-project defect prediction techniques from the literature. Specifically, we evaluate the conclusion stability of five defect prediction techniques that \citet{Herbold} recently evaluated in a defect prediction benchmarking study. We choose this study as a reference, because it is the most comprehensive evaluation of CPDP approaches, and evaluating techniques from their study allows us to compare our results with them. The results and replication kit of benchmarking study are also publicly available \citep{herbold2017repKit}.
	
	The five replicated techniques include the one proposed by \citet{Amasaki15} \textbf{(Amasaki15)}, \citet{Watanabe08} \textbf{(Watanabe08)},
	\citet{CamargoCruz09} \textbf{(CamargoCruz09)}, \citet{Nam15} \textbf{(Nam15)}, and \citet{Ma12} \textbf{(Ma12)}.
	The selection is guided by original rankings reported in the benchmarking study done by \citet{Herbold}.  CamargoCruz09 and Watanabe08 are the top-ranked techniques according to the rankings reported in \citet{Herbold}. The other two techniques, Amasaki15 and Ma12 are among the middle ranked approaches whereas Nam15 performs worst. Hence, to ensure diversity, we choose two top ranked, two middle ranked, and one lowest rank approach for evaluation.\footnote{In the rest of the paper, we do not use the rankings reported in original study of \citet{Herbold}, but instead  use our re-implementation results of his methodology on open-source projects in Jureczko data set.}
	
	We take a limited number of techniques, because of the large number of models that we already have to train at each point in time with varying window sizes. Our problem has a huge dimensionality and it could grow significantly by adding more techniques, because, for each new technique multiple Tr-Test pairs i.e. models need to be evaluated.
	
	\subsection{Extract software defect prediction data set with dated releases} To choose our data set, we explored the well-known PROMISE repository that is used in many defect-prediction studies \citep{DefectPredictionWork1, DefectPredictionWork2, DefectPredictionWork3, DefectPredictionWork4, DefectPredictionWork5}. Unfortunately, we could not find time-relevant features within that data set, which suggests the lack of concern about the time-order of defect data in the community. We also explored the five data sets used in the benchmarking study of \citet{Herbold}, but all except the Jureczko \citep{Jureckzo} lack time-relevant information that can be used to retrieve time of occurrence of defects. Since we need release-time information, we only use a subset of Jureczko data set consisting of only open-source projects, and we refer to it as \textsc{FilterJureczko}. We use open-source projects because their version numbers were specified, and hence release dates of only these versions could be retrieved from the project's version control repositories. As a result, we got 33~versions of 14~open-source projects for our experiment containing 20~static product metrics for Java classes and the number of defects found at class-level. Therefore our CPDP experiment is on class-level.
	
	\begin{figure}
		\centering
		\includegraphics[width=\columnwidth]{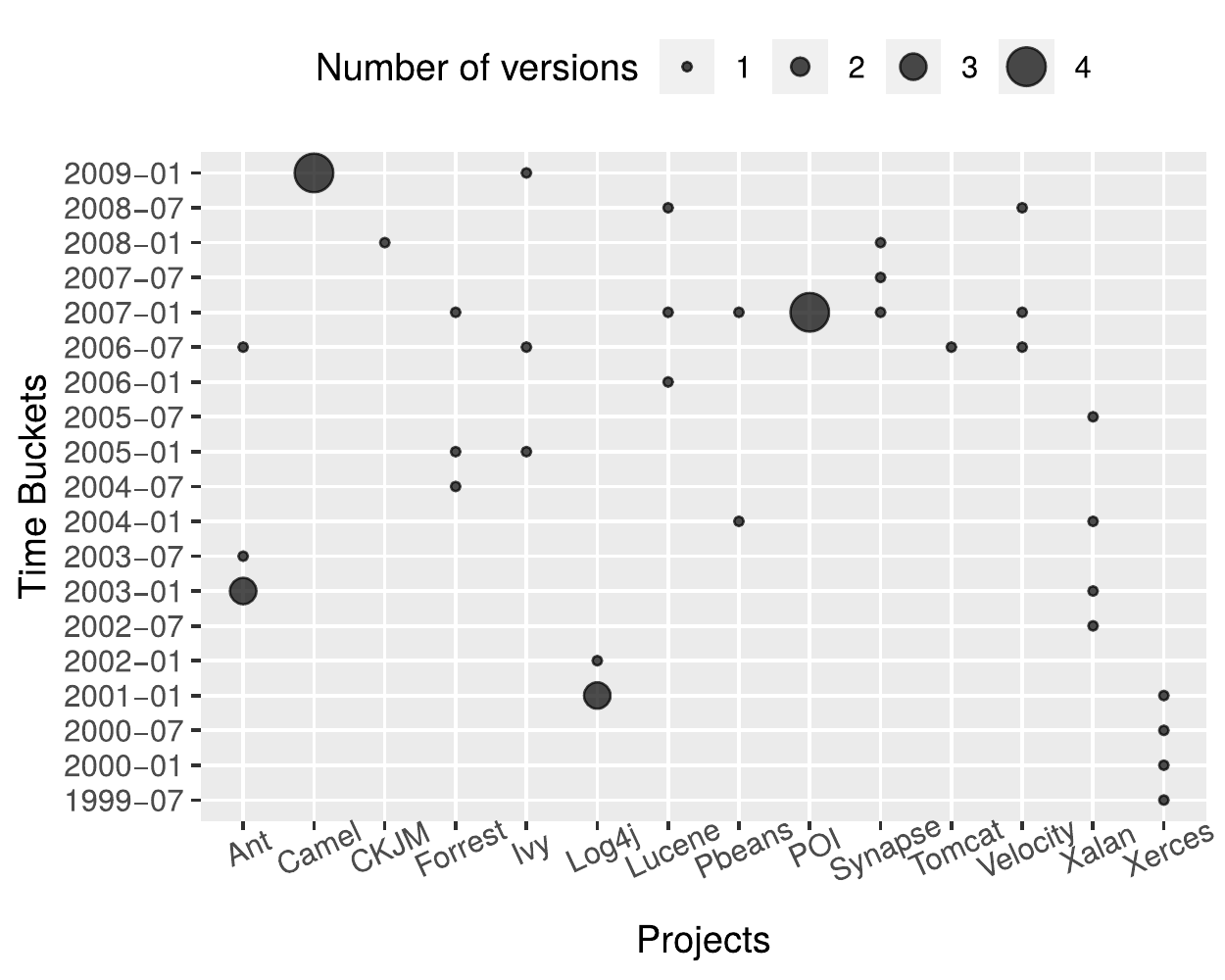}
		\caption{Project versions in our dataset spread across 19 time buckets. Number of projects represented by dot size corresponds to number of versions of a project in any time bucket shown on y-axis.}
		\label{fig:timeBuckets}
	\end{figure}
	
	\subsection{Sort and Split project versions into time buckets} The project versions in the \textsc{FilterJureczko} data set are spread across 8.5 years starting from November 1999 and ending at February 2009. We sort the entire data set using the version release dates and then divide it using split points having 6~month granularity. These points equally split the data set into a number of 6 month time-buckets; each containing project versions that are at most 6~months apart. We did not keep a finer granularity than 6 months, because of the limited data at hand and also because project releases are usually several months apart. In total, we have 18 buckets. Each bucket consists of multiple versions of different projects that lie within the 6-month time period.
	Out of 18 buckets, some buckets have multiple versions of the same project, because multiple versions were released within the 6-month time period whereas some buckets are completely empty because no project version was released during six months. In the end, we partitioned the entire data set into 18 sorted time-buckets and we refer to it as a ``time-series data set". \figref{fig:timeBuckets} is a graphical illustration of different project versions spread across 18 time buckets. For example, the first bucket has only one version of \textit{Xerces}, and the last bucket has four versions of \textit{Camel} and one version of \textit{Ivy}. \tabref{table:datasetInfo} represents the release date and defective instances for each version of the projects in our data set.
	
	\begin{table}
		\begin{center}
			\caption{the details of the \textsc{FilterJureczko} dataset, showing the version of each project with its release date and defective instances.}
			
			\begin{tabular}{l r r r r}
				\toprule
				\textbf{Version} & \textbf{Release Date} & \textbf{Cases} & \textbf{\#Defects} & \textbf{Defective Instances(\%)} \\ 
				\midrule
				
				xerces-init & 1999-Nov-08 & 162 & 77 & 48\% \\
				xerces-1.2 & 2000-Jun-23 & 440 & 71 & 16\% \\
				xerces-1.3 & 2000-Sep-29 & 453 & 69 & 15\% \\
				log4j-1.0 & 2001-Jan-08 & 135 & 34 & 25\% \\
				xerces-1.4 & 2001-Jan-26  & 588 & 437 & 74\% \\
				log4j-1.1 & 2001-May-20 & 109 & 37 & 34\% \\
				log4j-1.2 & 2002-May-10 & 205 & 189 & 92\% \\
				xalan-2.4 & 2002-Aug-28 & 723 & 110 & 15\% \\
				xalan-2.5 & 2003-Apr-10 & 803 & 387 & 48\% \\
				ant-1.3 & 2003-Aug-12 & 125 & 20 & 16\% \\
				ant-1.4 & 2003-Aug-12 & 178 & 40 & 22\% \\
				ant-1.5 & 2003-Aug-12 & 258 & 28 & 11\% \\
				ant-1.6 & 2003-Dec-18 & 351 & 92 & 26\% \\
				xalan-2.6 & 2004-Feb-27 & 885 & 411 & 46\% \\
				pbeans1.0 & 2004-Mar-21 & 26 & 20 & 77\% \\
				forrest-0.6 & 2004-Oct-14 & 6 & 1 & 17\% \\
				ivy-1.1 & 2005-Jun-13 & 111 & 63 & 57\% \\
				forrest-0.7 & 2005-Jun-22 & 29 & 5 & 17\% \\
				xalan-2.7 & 2005-Aug-06 & 909 & 898 & 99\% \\
				lucene-2.0 & 2006-May-26 & 195 & 91 & 47\% \\
				tomcat & 2006-Oct-21 & 858 & 77 & 9\% \\
				ivy-1.4 & 2006-Nov-09 & 241 & 16 & 7\% \\
				velocity-1.4 & 2006-Dec-01 & 196 & 147 & 75\% \\
				ant-1.7 & 2006-Dec-13 & 745 & 166 & 22\% \\
				velocity-1.5 & 2007-Mar-06 & 214 & 142 & 66\% \\
				pbeans2.0 & 2007-Mar-26 & 51 & 10 & 19\% \\
				forrest-0.8 & 2007-Apr-17 & 32 & 2 & 6\% \\
				synapse-1.0 & 2007-Jun-13 & 157 & 16 & 10\% \\
				lucene-2.2 & 2007-Jun-17 & 247 & 144 & 58\% \\
				poi-2.0 & 2007-Jun-24 & 314 & 37 & 12\% \\
				poi-1.5 & 2007-Jun-24 & 237 & 141 & 59\% \\
				poi-2.5 & 2007-Jun-24 & 385 & 248 & 64\% \\
				poi-3.0 & 2007-Jun-24 & 442 & 281 & 64\% \\
				synapse-1.1 & 2007-Nov-13 & 222 & 60 & 27\% \\
				synapse-1.2 & 2008-Jun-09 & 256 & 86 & 34\% \\
				ckjm1.8 & 2008-Jun-17 & 10 & 5 & 50\% \\
				lucene-2.4 & 2008-Oct-08 & 340 & 203 & 60\% \\
				velocity-1.6 & 2008-Dec-01 & 229 & 78 & 34\% \\
				ivy-2.0 & 2009-Jan-18 & 352 & 40 & 11\% \\
				camel-1.0 & 2009-Jan-19 & 339 & 13 & 4\% \\
				camel-1.2 & 2009-Jan-19 & 608 & 216 & 36\% \\
				camel-1.4 & 2009-Jan-19  & 872 & 145 & 17\% \\
				camel-1.6 & 2009-Feb-17 & 965 & 188 & 19\% \\
				
				\bottomrule
			\end{tabular}
			\label{table:datasetInfo}
		\end{center}
	\end{table}
	
	\subsection{Generate Train-Test pairs from time buckets}  
	%
	%
	%
	%
	%

	\begin{figure*}
		\centering
		\includegraphics[width=\textwidth]{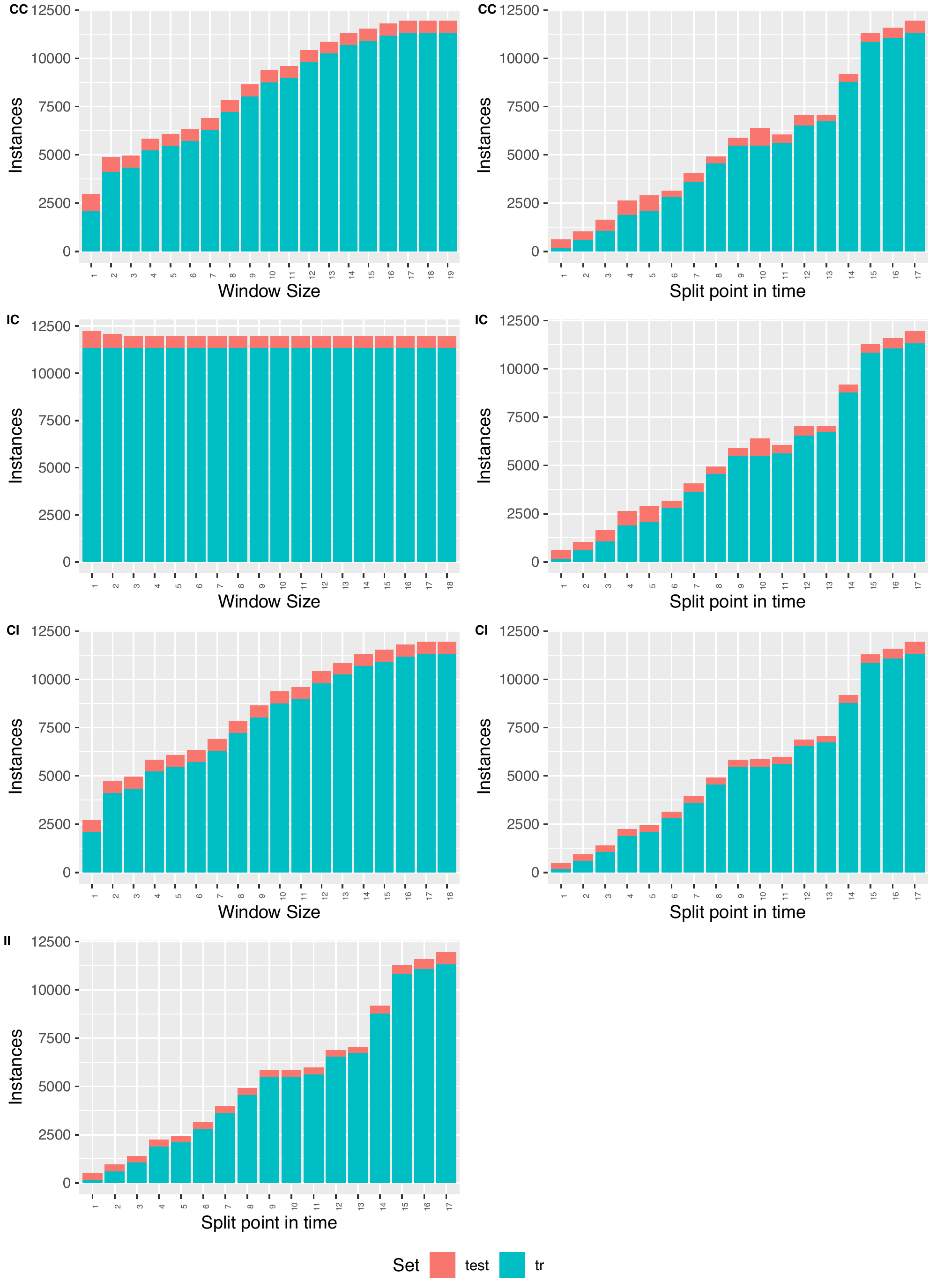}
		\caption{Representation of Training (tr) and Test (test) Data Set Size i.e. Number of instances with varying window size and split points in time for each configuration. Instances show the number of instances available in Train and Test set for n-th Window Size or n-th Split point in time. }
		\label{fig:TrainDataSize}
	\end{figure*}
	
	
	%
	%
	%
	
	\begin{figure*}
		\centering
		\includegraphics[width=\textwidth]{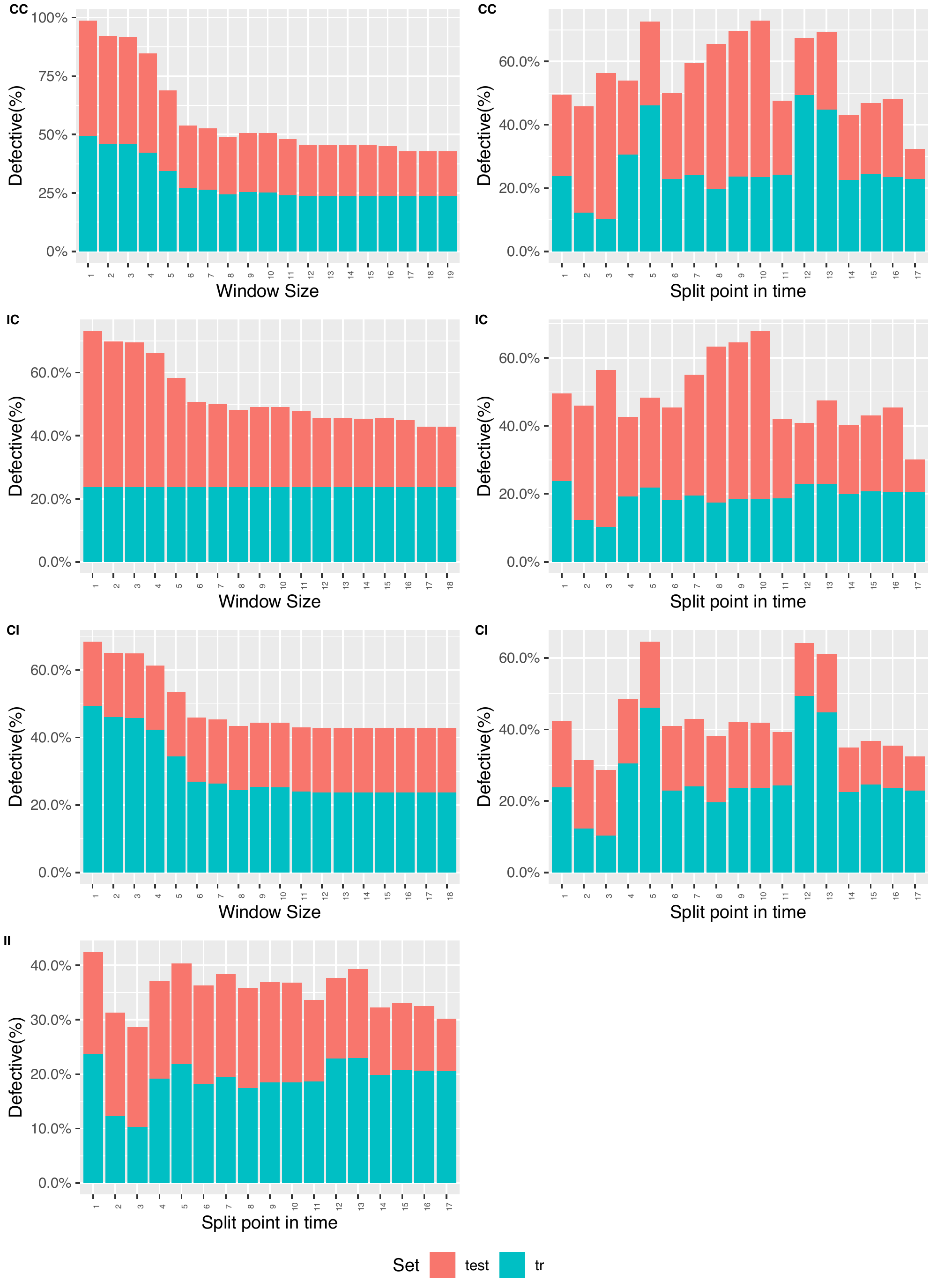}
		\caption{Representation of Defective Instances ($\%$) in Training (tr) and Test (test) Data Sets with varying window size and split points in time for each configuration.}
		\label{fig:Train_PercentDefects}
	\end{figure*}
	We generate multiple Tr-Test pairs from the time-series data set using four generic configurations; CC, IC, CI, and II. The Tr and Test sets are formed by varying the window size from 1 to 19 for CC and 1 to 18 for IC at all possible split points and then unioning the training project data. However, following strict CPDP, we do not allow the test set to include any version from a project that was already part of the training set. At the same time, to ensure that the data from which defect labels are computed does not intersect with test data from the next time bin, we leave a gap of one bucket between each Tr-Test pair, similar to the work
	by \citet{tan2015online}.

	We generated approximately 18,000~Tr-Test pairs for each technique and trained a total of $18,000\times5=90,000$ models for evaluation of the five techniques that we studied. The different number of Tr-Test pairs (and models) in CC, CI, and IC is due to the strict CPDP settings of our experiment, which does not allow the same project to be used for both training and testing. Consequently, at some split points, there is no data left for testing and hence we eliminate that pair.
	\figref{fig:TrainDataSize} shows the size of training and test data for each of the pairs in the four configurations. We also show the percentage of defective instances in our training and test data set at each split point and window size in \figref{fig:Train_PercentDefects}.

	\subsection{Build prediction models and evaluate performance} The defect prediction techniques apply certain treatments on the data before training the actual model. The treatments are applied as suggested by the benchmarking study of \citet{Herbold}. Suppose the training data is referred as $S$  and the test data is $S^*$.
	
	For \textbf{Amasaki15} \citep{Amasaki15}, we perform attribute selection over log transformed data by discarding attributes whose value is not close to any metric value in the data. We then apply relevancy filtering similarly by discarding instances whose value is not close to any instance values.
	
	For \textbf{Watanabe08} \citep{Watanabe08}, we standardize the training data for all Tr-Test pairs as:
	
	\begin{align*}
	\hat{m}_i(s^*)= (m_i(s^*) \cdot mean(m_i(S))) /  (mean(m_i(S^*)))
	\end{align*}
	
	For \textbf{CamargoCruz09} \citep{CamargoCruz09}, we use Test data as reference point and apply logarithmic transformation as:
	
	\begin{align*}
	\hat{m}_i(s)=  \log(1+m_i(s)) + median(log(1+m_i(S))) - median(log(1+m_i(S^*)))
	\end{align*}

	For \textbf{Nam15}  \citep{Nam15}, clustering and labelling of instances is performed based on the metric data by counting the number of attribute values that are above the median for that attribute. Afterwards all instances that do not violate a metric value based on a threshold called metric violation score are selected.
	
	For \textbf{Ma12} \citep{Ma12}, weighting is applied on data on the basis of similarity. The weights are calculated as:
	
	\begin{align*}
	w_s = simatts_s / (p-simmats_s + 1)^2
	\end{align*}
	
	where $p$ is the number of attributes and $simatts$ are those attributes of an instance whose value is within the range of test data.
	
	More details about these techniques are available in their original publications. The source code for applying these treatments is provided by \citet{crosspare2015,herbold2017repKit} as a replication package.\footnote{Herbold's replication kit (https://crosspare.informatik.uni-goettingen.de/)}
	
	For each technique, we built 976~separate defect prediction models utilizing all the Tr-Test pairs. We trained these models on Decision Trees (DT) using C4.5 algorithm in Weka \citep{weka}. We chose DT, because all the studied techniques performed best on Decision Trees classifier in the benchmarking study \citep{Herbold}. To compare our results with the benchmarking study, we also trained our models on DT using a confidence interval of between 0.1 and 0.30~with pruning. We did not tune our classifier to keep the experimental settings consistent with \citet{Herbold}, because changing them could bias our results and the observed difference in performance could entirely be due to tuning. Moreover, our small data set limits us from giving up a whole window for tuning. \citet{rakha2018} had an ample amount of data, hence they tuned their models in the duplicate issue reports study.
	
	While evaluating our models, we calculated their performance in terms of precision, recall, F-Score, G-measure, MCC, and AUC.
	\textit{Recall} is the ratio of true positives to true positives and false negatives, and it measures the number of actual defects that are found. \textit{Precision} is the ratio of true positives to true positives and false positives, and it measures how many of the found defects are actually defects. \textit{F-Score} is a combination of precision and recall, and is calculated using the harmonic mean of the two. \textit{G-measure} is the harmonic mean of recall and the probability of false prediction, \textit{pf}. \textit{Matthews Correlation Coefficient (MCC)} measures the correlation between the actual and the predicted classifications, ranging between -1 and +1, where -1 indicates total disagreement, +1 indicates perfect agreement, and 0 indicates no correlation at all. \textit{AUC} or the Area under the Receiver Operating Characteristic Curve is a plot of the true positive rate vs the true negative rate. These performance metrics are defined as follows, 
	
	\begin{equation*}
	recall= \frac{tp}{tp+fn}
	\label{recallEq}
	\end{equation*}

	\begin{equation*}
	precision= \frac{tp}{tp+fp}
	\label{precisionEq}
	\end{equation*}

	\begin{equation*}
	F-score = 2\cdot \frac{precision\cdot recall} {recall + precision}
	\label{fscoreEq}
	\end{equation*}

	\begin{equation*}
	\begin{split}
	G-measure = 2\cdot\frac{recall \cdot(1-pf)} {recall + (1-pf)}   \\ where,\  pf=\frac{fp}{tn+fp}
	\end{split}
	\label{gscoreEq}
	\end{equation*}

	\begin{equation*}
	MCC =\frac{ tp\cdot tn - fp\cdot fn}{\sqrt{(tp+fp)(tp+fn)(tn+fp)(tn+fn)}}
	\label{mccEq}
	\end{equation*}
	
	where \textit{tp} and \textit{fp} are the numbers of the true and false positives respectively, whereas, \textit{tn} and \textit{fn} are the numbers of the true and false negatives.
	\begin{table}
		\begin{center}
			\caption{ Comparison of our methodology and experimental setup with the original study of \citet{Herbold} and \textsc{HerboldMethod} which is our re-implementation of their study}
			
			\begin{tabular}{l p{8em} p{8em} p{8em}}
				\toprule
				\textbf{Evaluation\ Parameter} & \textbf{Original Study \citep{Herbold}} & \textsc{\textbf{HerboldMethod}}& \textbf{Time-aware Evaluation} \\ \midrule
				
				CPDP Type & Strict  &	Strict & Strict \\
				
				\cmidrule(lr){1-4}
				
				Approaches Evaluated & 24 & 5 & 5 \\
				
				\cmidrule(lr){1-4}						
				
				Datasets &	Jureczko and three others & \textsc{FilterJureczko}  & \textsc{FilterJureczko}\\
				
				\cmidrule(lr){1-4}
				
				Data time considered  & No & No & Yes\\
				
				\cmidrule(lr){1-4}
				
				Classifiers & Decision tree and five more & Decision tree &	Decision tree  \\
				\cmidrule(lr){1-4}
				
				Data balancing & No & No & No\\
				\cmidrule(lr){1-4}
				
				Classifier Tuning & No & No & No\\
				\cmidrule(lr){1-4}
				
				Classifier Training  & Cross-validation & Cross-validation & Four time-aware configurations \\
				\cmidrule(lr){1-4}
				
				Performance Metrics & {F-measure, MCC, AUC, G-measure, Mean-rank score} &  {F-measure, MCC, AUC, G-measure, Mean-rank score} &
				{F-measure, MCC, AUC, G-measure, Mean-rank score}\\
				
				\bottomrule
			\end{tabular}
			\label{table:methodDiff}
		\end{center}
	\end{table}
	
	\section{Results}
	As a result of running our time-aware experiment we gather models for each Tr-Test pair representing one split point in time and each window size of a given configuration. All the models are built using Decision Tree classifier and the results constitute a range of performance estimates that we use to examine conclusion stability of cross-project defect prediction models. We also compare the results of our time-aware experiment with results obtained by re-conducting the experiment of \citet{Herbold} on the \textsc{FilterJureczko} data set. Instead of reporting the result of Herbold's original study, we use our re-implementation results of his methodology referred subsequently as \textsc{HerboldMethod}. \tabref{table:methodDiff} highlights some of the commonalities and differences between our evaluation and \textsc{HerboldMethod}. However, different research questions can also be answered using our methodology. To facilitate further investigations, we provide a replication kit \citep{aliEMSE2020} which includes:
	
	\begin{enumerate}
		\item \textsc{FilteredJureczko} data set with Tr-Test pairs of all four configurations.
		\item a source code for generating four configurations Tr-Test pairs from any data set.
		\item an updated version of Herbold's source code for time-aware experiment.
		\item a .csv dump file for all the results calculated from our experiment.
		\item R-scripts to generate graphs for visual inspection of results.
	\end{enumerate}
	\textit{Replication kit: https://doi.org/10.5281/zenodo.3715485}

	\subsection{\textbf{RQ1: Are the cross-project defect prediction approaches stable in terms of their conclusions when evaluated over  time?}}
	
	\begin{figure*}
		\centering
		\includegraphics[width=\textwidth]{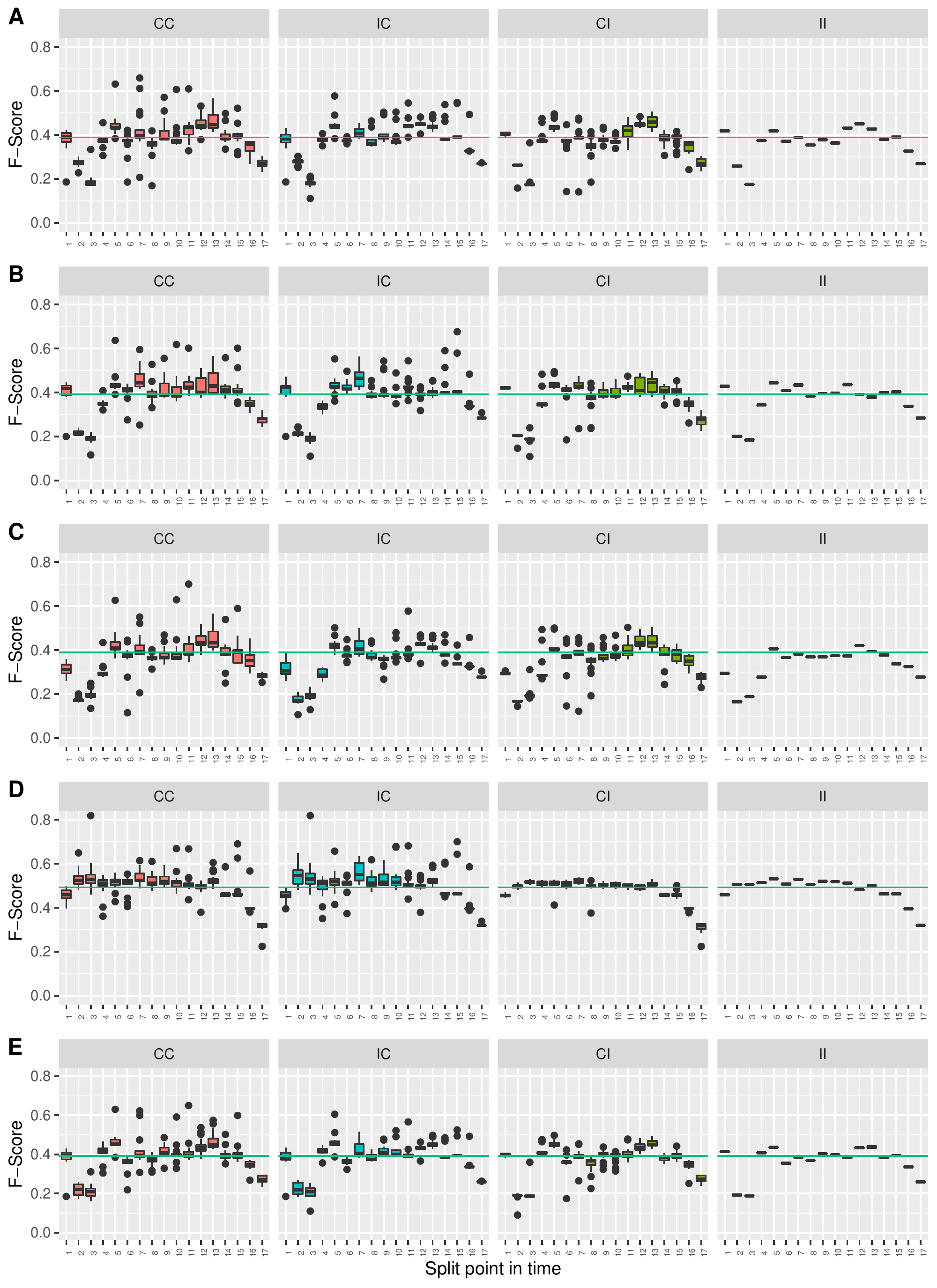}
		\caption{Comparison of F-Scores of techniques when evaluated over four configurations. A-Amasaki15, B-Watanabe08, C-CamargoCruz09, D-Nam15, E-Ma12. Horizontal line shows \textsc{HerboldMethod} F-Score}
		\label{fig:techniquesComp}
	\end{figure*}
	\begin{table}
		\centering
		\caption{Arithmetic Mean and Standard Deviation(SD) of the F-Scores of five evaluated approaches using four time-aware configurations. Bold values indicate SD larger than our 0.05 threshold}
		\resizebox{\textwidth}{!}{
			\begin{tabular}{ccccccccccc}
				\toprule
				\multicolumn{1}{c}{} & \multicolumn{2}{c}{\bf Amasaki15} & \multicolumn{2}{c}{\bf Watanabe08}& \multicolumn{2}{c}{\bf CamargoCruz09}& \multicolumn{2}{c}{\bf Nam15}& \multicolumn{2}{c}{\bf Ma12}\\

				\bf Configuration & Mean& SD &  Mean& SD  & Mean& SD & Mean& SD & Mean&SD \\
				
				\cmidrule(lr){1-1} \cmidrule(lr){2-3} \cmidrule(lr){4-5} \cmidrule(lr){6-7} \cmidrule(lr){8-9} \cmidrule(lr){10-11}
				
				CC & 0.373 & \textbf{0.085} & 0.379 & \textbf{0.089} & 0.355 & \textbf{0.091} & 0.491 & \textbf{0.070} & 0.376 & \textbf{0.085} \\ 
				IC & 0.373 & \textbf{0.078} & 0.373 & \textbf{0.084} & 0.345 & \textbf{0.079} & 0.496 & \textbf{0.072} & 0.376 & \textbf{0.081} \\ 
				CI & 0.365 & \textbf{0.077} & 0.371 & \textbf{0.083} & 0.346 & \textbf{0.084} & 0.478 & \textbf{0.055} & 0.366 & \textbf{0.082} \\ 
				II & 0.363 & \textbf{0.071} & 0.367 & \textbf{0.077} & 0.335 & \textbf{0.073} & 0.483 & \textbf{0.054} & 0.363 & \textbf{0.078} \\

				\hline
		\end{tabular}}
		\label{table:meanAndSdFscore}
	\end{table}
	\begin{table}
		\centering
		\caption{Arithmetic Mean and Standard Deviation(SD) of the AUCs of five evaluated approaches using four time-aware configurations.}
		\resizebox{\textwidth}{!}{
			\begin{tabular}{ccccccccccc}
				\toprule
				\multicolumn{1}{c}{} & \multicolumn{2}{c}{\bf Amasaki15} & \multicolumn{2}{c}{\bf Watanabe08}& \multicolumn{2}{c}{\bf CamargoCruz09}& \multicolumn{2}{c}{\bf Nam15}& \multicolumn{2}{c}{\bf Ma12}\\

				\bf Configuration & Mean& SD &  Mean& SD  & Mean& SD & Mean& SD & Mean&SD \\
				
				\cmidrule(lr){1-1} \cmidrule(lr){2-3} \cmidrule(lr){4-5} \cmidrule(lr){6-7} \cmidrule(lr){8-9} \cmidrule(lr){10-11}

				CC&	0.571 &	0.045 &	0.562 &	0.046 &	0.556 &	0.042 &	0.638&	0.021&	0.577&	0.036\\
				IC&	0.573 &	0.038 &	0.555 &	0.044 &	0.548 &	0.039 &	0.638&	0.021&	0.579&	0.029\\
				CI&	0.572 &	0.036 &	0.562 &	0.040 &	0.559 &	0.038 &	0.636&	0.015&	0.578&	0.028\\
				II&	0.571 &	0.034 &	0.557 &	0.038 &	0.549 &	0.034 &	0.636&	0.014&	0.577&	0.024\\

				\hline
		\end{tabular}}
		\label{table:meanAndSdAUC}
	\end{table}
	\begin{table}
		\centering
		\caption{Arithmetic Mean and Standard Deviation(SD) of the MCCs of five evaluated approaches using four time-aware configurations. Bold values indicate SD is larger than our 0.05 threshold}
		\resizebox{\textwidth}{!}{
			\begin{tabular}{ccccccccccc}
				\toprule
				\multicolumn{1}{c}{} & \multicolumn{2}{c}{\bf Amasaki15} & \multicolumn{2}{c}{\bf Watanabe08}& \multicolumn{2}{c}{\bf CamargoCruz09}& \multicolumn{2}{c}{\bf Nam15}& \multicolumn{2}{c}{\bf Ma12}\\

				\bf Configuration & Mean& SD &  Mean& SD  & Mean& SD & Mean& SD & Mean&SD \\
				
				\cmidrule(lr){1-1} \cmidrule(lr){2-3} \cmidrule(lr){4-5} \cmidrule(lr){6-7} \cmidrule(lr){8-9} \cmidrule(lr){10-11}
				
				CC& 0.143&	\textbf{0.055}&	0.124 &	\textbf{0.061} & 0.117&	\textbf{0.064} &	0.232 &	0.037&	0.135&	\textbf{0.051}\\
				IC& 0.156&	\textbf{0.053}&	0.127 &	\textbf{0.063} & 0.118&	\textbf{0.060} &	0.231 &	0.040&	0.142&	0.046\\
				CI& 0.141&       	0.045 &	0.125 &	\textbf{0.054} & 0.116&	\textbf{0.055} &	0.228 &	0.026&	0.134&	0.043\\
				II& 0.152&	        0.045 &	0.131 &	\textbf{0.050} & 0.114&	\textbf{0.055} &	0.229 &	0.027&	0.140&	0.044\\

				\hline
		\end{tabular}}
		\label{table:meanAndSdMCC}
	\end{table}
	\begin{table}
		\centering
		\caption{Arithmetic Mean and Standard Deviation(SD) of the G-measures of five evaluated approaches using four time-aware configurations. Bold values indicate SD is larger than our 0.05 threshold}
		\resizebox{\textwidth}{!}{
			\begin{tabular}{ccccccccccc}
				\toprule
				\multicolumn{1}{c}{} & \multicolumn{2}{c}{\bf Amasaki15} & \multicolumn{2}{c}{\bf Watanabe08}& \multicolumn{2}{c}{\bf CamargoCruz09}& \multicolumn{2}{c}{\bf Nam15}& \multicolumn{2}{c}{\bf Ma12}\\

				\bf Configuration & Mean& SD &  Mean& SD  & Mean& SD & Mean& SD & Mean&SD \\
				
				\cmidrule(lr){1-1} \cmidrule(lr){2-3} \cmidrule(lr){4-5} \cmidrule(lr){6-7} \cmidrule(lr){8-9} \cmidrule(lr){10-11}

				CC& 0.483&	\textbf{0.113} & 0.497 & \textbf{0.121} & 0.464 & \textbf{0.121} & 0.582 & \textbf{0.058} &	0.491 &	\textbf{0.111}\\
				IC& 0.498&	\textbf{0.092} & 0.510 & \textbf{0.103} & 0.471 & \textbf{0.103} & 0.581 & \textbf{0.054} &	0.506 &	\textbf{0.093}\\
				CI& 0.482&	\textbf{0.112} & 0.497 & \textbf{0.123} & 0.464 & \textbf{0.119} & 0.586 &         0.048  &	0.485 &	\textbf{0.116}\\
				II& 0.494&	\textbf{0.093} & 0.513 & \textbf{0.104} & 0.470 & \textbf{0.102} & 0.585 &         0.041  &	0.501 &	\textbf{0.099}\\

				\hline
		\end{tabular}}
		\label{table:meanAndSdGscore}
	\end{table}
	
	\paragraph{Motivation.}  Prior research evaluates defect prediction approaches in a time-agnostic manner. The results obtained from one specific evaluation at a particular point in time are generalized to all available time-periods. This assumption is unrealistic as defect prediction approaches might not have stable conclusions and hence results cannot be generalized across the entire data set irrespective of time. The goal of this research question is to study the conclusion stability of defect prediction approaches. We hypothesize that ``a defect prediction technique has stable conclusion if for a given performance metric, the standard deviation produced by all Tr-Test pairs in a specific configuration is less than absolute 0.05". Prior works such as \citet{zhang2014} and \citet{Herbold} consider 2\% and 5\% respectively to be a significant performance gain in terms of AUC and F-Score, therefore we also use 0.05 absolute value of a performance metric for the threshold. 

	\paragraph{Result.} We evaluate five cross-project defect prediction approaches in this paper and according to the results of our experiment these approaches have unstable conclusions. To investigate the conclusion stability; we analyze the F-Score, AUC, MCC and G-measure values obtained from different evaluations of five approaches using Tr-Test pairs generated according to the four configurations introduced earlier. 
	\tabref{table:meanAndSdFscore} through 
	\tabref{table:meanAndSdGscore} shows the mean and the standard deviation of F-Score, AUC, MCC, and G-measure for the five evaluated approaches. The mean and standard deviation values were calculated across all Tr-Test pairs generated according to CC, CI, IC and II configuration.  
	
	The bold values in \tabref{table:meanAndSdFscore} through \tabref{table:meanAndSdGscore} indicate that the overall standard deviation of the given performance metric observed across different evaluations in a configuration is greater than 0.05. The F-Scores and the G-measures of all five approaches vary by more than 0.05 in almost all configurations suggesting that instability exists. We believe that conclusions of a model's performance may change depending on the context, i.e., time at which model was trained and evaluated which explains this instability in all performance metrics except AUC which remains stable. This conclusion about the performance of models with respect to time is re-assured in \secref{sec:constantWindowSize} by measuring the F-Score standard deviation while keeping the window size constant.
	
	\figref{fig:techniquesComp} further shows F-Scores plotted on y-axis over split points in time on the x-axis. The boxplots in figure illustrate the variance in the F-Score values of techniques evaluated according to four configurations. The length of barplots signify the magnitude of variation in the F-Score at a particular split point. If we observe the F-Score values along the timeline in \figref{fig:techniquesComp}, there is a drastic variation at different points in time, particularly for CC and IC and to a relatively lesser extent in CI. In the II configuration, the F-Scores of all techniques except Nam15 exhibit a similar variation across timeline. Overall CamargoCruz09 shows the highest deviation by deviating more than 0.05 from it's mean value almost 26\% of the times followed by Amasaki15, Watanabe08, Ma12 and Nam15 respectively which deviate 25\%, 24\%, 24\% and 11\% of the times respectively.
	
	Since the time-agnostic evaluation ignores time, therefore all prior works report aggregate F-Score over the entire evaluated time-period. The green constant horizontal line drawn over \figref{fig:techniquesComp} refers to the F-Score value obtained by \textsc{HerboldMethod} and represents the mean of cross-validation F-Scores produced in different folds. The large number of results falling on both sides of the horizontal line indicate that conclusions drawn about the performance of an approach are not stable over different evaluations. For example, at split point 3 in CC configuration in \figref{fig:techniquesComp}-D, F-Score is above 0.8 but it drops to around 0.35 if we move just one split point ahead on the timeline to split 4. Such abrupt variations across the time line show that performance claims can be highly unrealistic if the context is ignored. Therefore reporting a single value and generalizing it over different points in a project's evolution can be quite misleading. 
	
	The problem is further aggravated by large number of outliers that can be seen in \figref{fig:techniquesComp}, indicating the fact that evaluation can often yield very high or low performance estimates, which are far from the real performance that a defect prediction technique may achieve in practice. Therefore, the conclusions drawn from a specific period of time should not be generalized outside of it. It is rather more appropriate for researchers to report a range of values of a performance metric corresponding to multiple time-periods and contexts of evaluation.

	\roundbox{
		The defect prediction techniques do not have stable conclusions when evaluated over several different points in time using four configurations. The G-measures of all techniques except Nam15 deviate more than 0.1 from their mean values in all configurations. Similarly F-Scores and MCCs deviate by 0.05 in at least one configuration. This deviation in performance metrics signifies that the performance based on one evaluated period of time cannot be generalized across the entire project or data set irrespective of time. Researchers should carefully couch the results of defect prediction studies against the time-periods of evaluation.}
	
	
	\subsection{ \textbf{RQ2: How do the results of time-agnostic and time-aware evaluations differ? }}
	
	\begin{table}
		\begin{center}
			\vspace{3em}
			\caption{Resulting \textit{p-values} of Wilcoxon rank-sum tests for comparison between four configurations and \textsc{HerboldMethod} for the five approaches. Bold values indicate statistically significant differences at $\alpha$ = 0.01 }
			\begin{tabular}{lrrrr}
				\toprule
				
				\bf Technique  & \textbf{F-Score} & \textbf{AUC} &  \textbf{MCC} & \textbf{G-measure}  \\
				
				\cmidrule(lr){1-1}  \cmidrule(lr){2-5}

				Amasaki15 &	\textbf{$<$ 0.01} & 0.20  & \textbf{$<$ 0.01} & 0.79 \\

				CamargoCruz09 & \textbf{$<$ 0.01} & \textbf{$<$ 0.01} & \textbf{$<$ 0.01} & \textbf{$<$ 0.01}\\

				Watanabe08	&  0.17     &  0.12		& \textbf{$<$ 0.01}  & \textbf{$<$ 0.01} \\

				Nam15  &   \textbf{$<$ 0.01} &  0.07 &  0.16 & \textbf{$<$ 0.01} \\				 
				
				Ma12 &      0.73 &  \textbf{$<$ 0.01} & \textbf{$<$ 0.01} & \textbf{$<$ 0.01}\\

				\hline
			\end{tabular}
			\label{table:wilcox}
		\end{center}
	\end{table}
	
	\paragraph{Motivation.} The time-agnostic evaluation of defect prediction techniques might lead to false estimates of performance. In this question we compare the results of time-agnostic and time-aware evaluations to better understand the impact of evaluation method on the results of cross-project defect prediction models.
	
	\paragraph{Result.} We use Wilcoxon rank-sum test to evaluate whether the differences between \textsc{HerboldMethod} and our results are statistically significant or not. \tabref{table:wilcox} reports the p-values of Wilcoxon test and bold values indicate a statistically significant difference at an $\alpha$ value of 0.01. The comparison reveals that the results of our time-aware evaluations differ from \textsc{HerboldMethod} in terms of all four metircs for the CamargoCruz09 and in terms of at least two out of the four metrics for the remaining approaches. These difference are also statistically significant (p-value $<$ 0.01).
	
	To quantify the differences between our configurations and \textsc{HerboldMethod} we employ \textit{Cliff’s Delta} which is a measure of the effect size and does not assume normality of distribution. For \textit{Cliff’s Delta} we use the interpretations of \citet{romano2006Cliff} which considers difference to be Negligible if Cliff’s $|d|$ $\leq$ 0.147, Small if Cliff’s $|d|$ $\leq$ 0.33, Medium when Cliff’s $|d|$ $\leq$ 0.474, and Large otherwise. The Cliff delta indicates that the observed differences have small to negligible effect size for all four metrics and five approaches. Despite an overall small effect size, the variations in performance at different split points and the combined effect of variations across different metrics cannot be ignored. Furthermore, regardless of the effect size, it is methodologically incorrect to evaluate the defect prediction techniques using time-agnostic evaluation or to generalize their performance beyond the evaluated time periods.
	
	\roundbox{
		The Wilcoxon rank-sum tests suggest that there is a statistically significant difference between the time-aware experiment and \textsc{HerboldMethod} on the basis of F-Score, AUC, MCC and G-measure but the \textit{Cliff's Delta} effect sizes are small to negligible.}
	
	\begin{table}
		\centering
		\caption{Raw result values of \textsc{HerboldMethod} and time-aware evaluation---\textsc{HerboldMethod} reports only one value of F-Score, AUC and MCC for each technique which is duplicated across all rows.}
		\resizebox{\textwidth}{!}{
			\begin{tabular}{llrrrrrrrr}
				\toprule
				&	& \multicolumn{4}{c}{\bf New Values} & \multicolumn{4}{c}{\bf \textsc{HerboldMethod} Values}\\
				
				\cmidrule{3-6}	\cmidrule(lr){7-10}
				
				\bf Technique & \bf Configuration & \textbf{F-Score} & \textbf{MCC} &  \textbf{AUC} & \textbf{G-Measure} & \textbf{F-Score} & \textbf{MCC} & \textbf{AUC} & \textbf{G-Measure} \\
				
				\cmidrule(lr){1-1} \cmidrule(lr){2-2}
				\cmidrule(lr){3-6} 	\cmidrule(lr){7-10}
				
				Amasaki15 & CC & 0.373 & 0.142 & 0.571 & 0.483 
				&   0.388 &    0.175 &        0.578  & 0.516 \\
				& IC &  0.373 &    0.155  &    0.573  & 0.498 
				&   0.388 &    0.175 &        0.578  & 0.516  \\
				& CI  & 0.365  & 	0.141  & 0.571  & 0.482 
				&   0.388 &    0.175 &        0.578  & 0.516  \\
				& II  & 0.363 & 0.152  & 0.571 & 0.494 
				&   0.388 &    0.175 &        0.578   & 0.516 \\
				& All Configurations &  0.368 &    0.148  &    0.571  & 0.489 
				&   0.388 &    0.175 &        0.578  & 0.516  \\

				\midrule

				Watanabe08 &  CC & 0.379 & 	0.123 & 0.562 & 0.497 
				&  0.392 & 0.109 & 0.563  & 0.506 \\
				&  IC &  0.373 & 	0.127 & 	0.555 & 0.510 
				&  0.392 &     0.109  &     0.563   & 0.506 \\
				&  CI &  0.371 & 	0.124 & 	0.562 & 0.497 
				&  0.392 &     0.109  &     0.563  & 0.506  \\
				&  II & 0.367 & 	0.131 & 	0.557 & 0.513 
				&  0.392 &     0.109  &     0.563  & 0.506  \\
				&  All Configurations & 0.373 &	0.126 & 0.559 & 0.504 
				&  0.392 &     0.109  &     0.563   & 0.506 \\

				\midrule
				
				CamargoCruz09 & CC &  0.354 & 0.117 & 0.556  & 0.464 
				&    0.389 &        -0.086 &    0.468 & 0.523  \\
				& IC & 0.345 &	0.118 & 0.548  & 0.471 
				&    0.389 &       -0.086 &         0.468 & 0.523 \\
				& CI &  0.346 & 0.116 & 0.559 & 0.464 
				&    0.389 &       -0.086 &         0.468 & 0.523 \\
				& II &  0.335 &	0.114 &	0.549 & 0.470 
				&    0.389 &       -0.086 &         0.468 & 0.523 \\
				& All Configurations &  0.345 &	0.116 &	0.553 & 0.467 
				&    0.389 &       -0.086 &         0.468 & 0.523 \\

				\midrule
				
				Nam15 &   CC  & 0.491 & 0.232  & 0.638    & 0.582 
				&    0.492 & 0.235 & 0.641 & 0.602 \\
				&   IC & 0.496 & 	0.231 & 	0.638 & 0.581 
				&    0.492 &       0.235 &        0.641 & 0.602  \\
				&   CI & 0.478 &       0.228 &        0.636 & 0.585 
				&    0.492 &       0.235 &        0.641  & 0.602 \\
				&   II & 0.483 &       0.229 &        0.636 & 0.585 
				&    0.492 &       0.235 &        0.641  & 0.602 \\
				&   All Configurations & 0.487 &   0.230 &   0.637 & 0.583 
				&    0.492 &       0.235 &        0.641  & 0.602 \\

				\midrule
				
				Ma12 &    CC & 0.376 &	0.134 &	0.577  & 0.491 
				&    0.392 & 0.160 & 0.581   & 0.521  \\
				&    IC & 0.376 &	0.142 &	0.578 & 0.506 
				&     0.392 &       0.160  &       0.581   & 0.521  \\
				&    CI & 0.366 &	0.134 & 0.577  & 0.485 
				&     0.392 &       0.160  &       0.581  & 0.521   \\
				&    II & 0.363 &	0.140 &	0.577  & 0.501 
				&     0.392 &       0.160  &       0.581  & 0.521   \\
				&    All Configurations & 0.370 &	0.138 &	0.577  & 0.496 
				&     0.392 &       0.160  &       0.581   & 0.521  \\
				\bottomrule

		\end{tabular}}
		\label{table:newTechniqueValues}    
	\end{table}

	\subsection{\textbf{RQ3: What is the ranking of evaluated techniques in time-aware experiment?}}
	
	\paragraph{Motivation.} The recent replication done by \citet{Herbold} ranks 24 cross project defect prediction approaches using common data sets and performance metrics. The aim of their work was to benchmark the performance of CPDP approaches using multiple learners and data sets. We on the other hand claim and show that their conclusion might not hold under different contexts of evaluation. In this research question, we investigate if the rankings of \textsc{HerboldMethod} still holds under our experimental settings or not.
	
	\paragraph{Result.} The performance estimates of our four configurations in comparison with \textsc{HerboldMethod} are reported in \tabref{table:newTechniqueValues}. The prior analysis in
	RQ2 suggests that for all approaches, the performance in terms of at least two evaluation metrics differ significantly.
	We further re-rank the defect prediction techniques relative to others on the basis of each performance metric, using the following formula suggested by \citet{Herbold}
	
	\begin{equation*}
	rankscore = 1- {\frac{\#{approaches\ ranked\ higher}} {\#{approaches - 1}} }
	\label{rankscore}
	\end{equation*} 
	
	The \textit{rankscore} lies in the range of 0 and 1 which respectively represents lowest and highest possible ranks. The \textit{Mean Rank Score} of a technique is the arithematic mean of rankscores computed using each of the four performance metric. Ranking a technique using all performance metrics reduces the bias arising due to a single metric failing to estimate the model performance. As a result, two approaches achieving same rankscore using two different metrics may have the same overall score. The ranks of each technique per configuration and the \textsc{HerboldMethod} ranks are presented in \tabref{table:newRanks}. Note that these ranks were calculated using \textit{Mean Rank Score} but, for the sake of readability, the decimal values of \textit{Mean Rank Score} were replaced with the respective ranks that those values represent.
	
\begin{figure*}
		\centering
		\includegraphics[width=\textwidth]{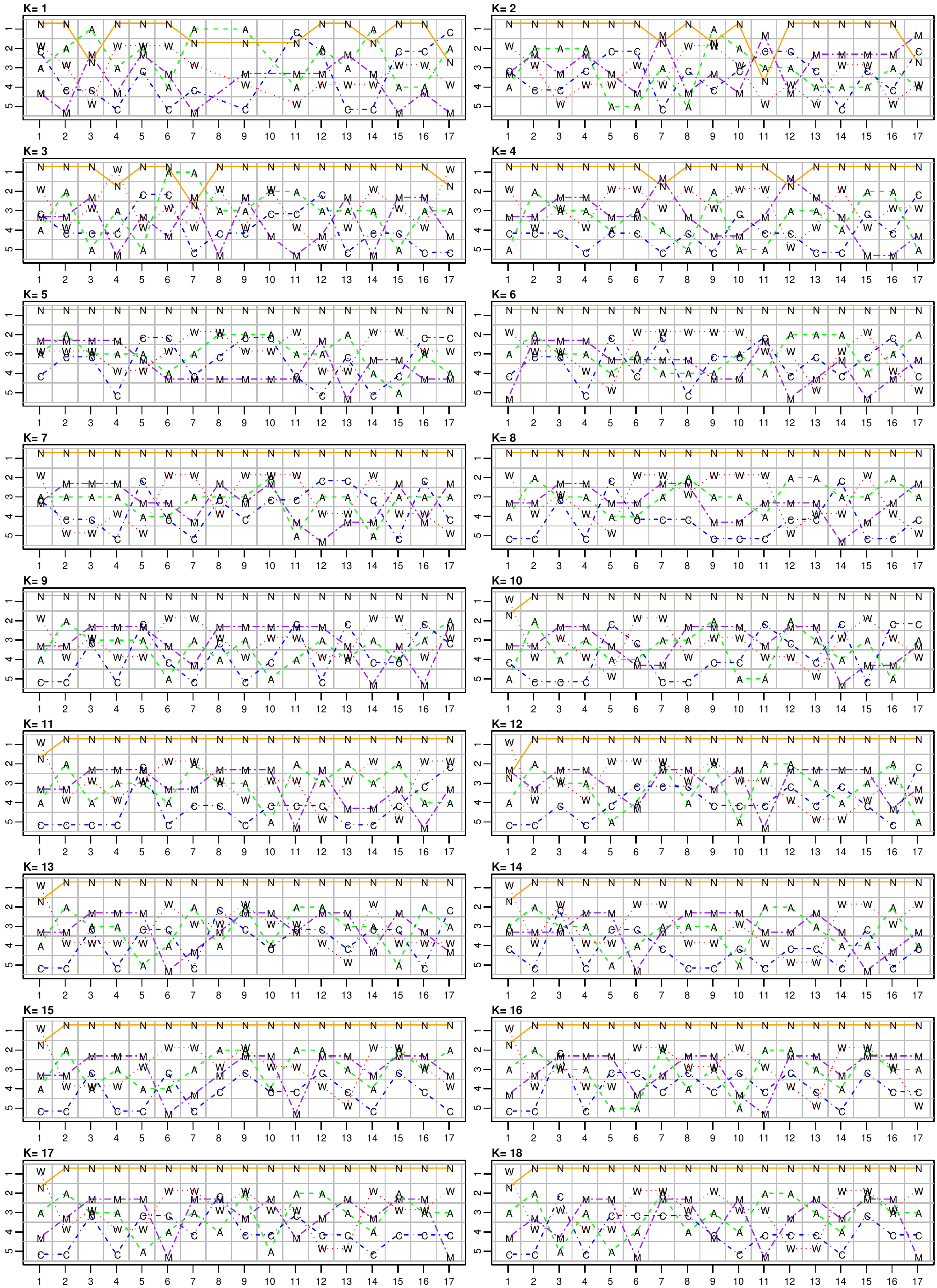}
		\caption{Variation in ranks of techniques evaluated using CC configuration. Each sub-figure represents a window size from (1 to 17), x-axis shows split point in time (1 to 17), y-axis shows the ranks of technique from (1 to 5), and K represents window size. Techniques: A=Amasaki15, W=Watanabe08, C=CamargoCruz09, N=Nam15, M=Ma12.}
		
		\label{fig:stepConfigCC}
	\end{figure*}

\begin{figure*}
		\centering
		\includegraphics[width=\textwidth]{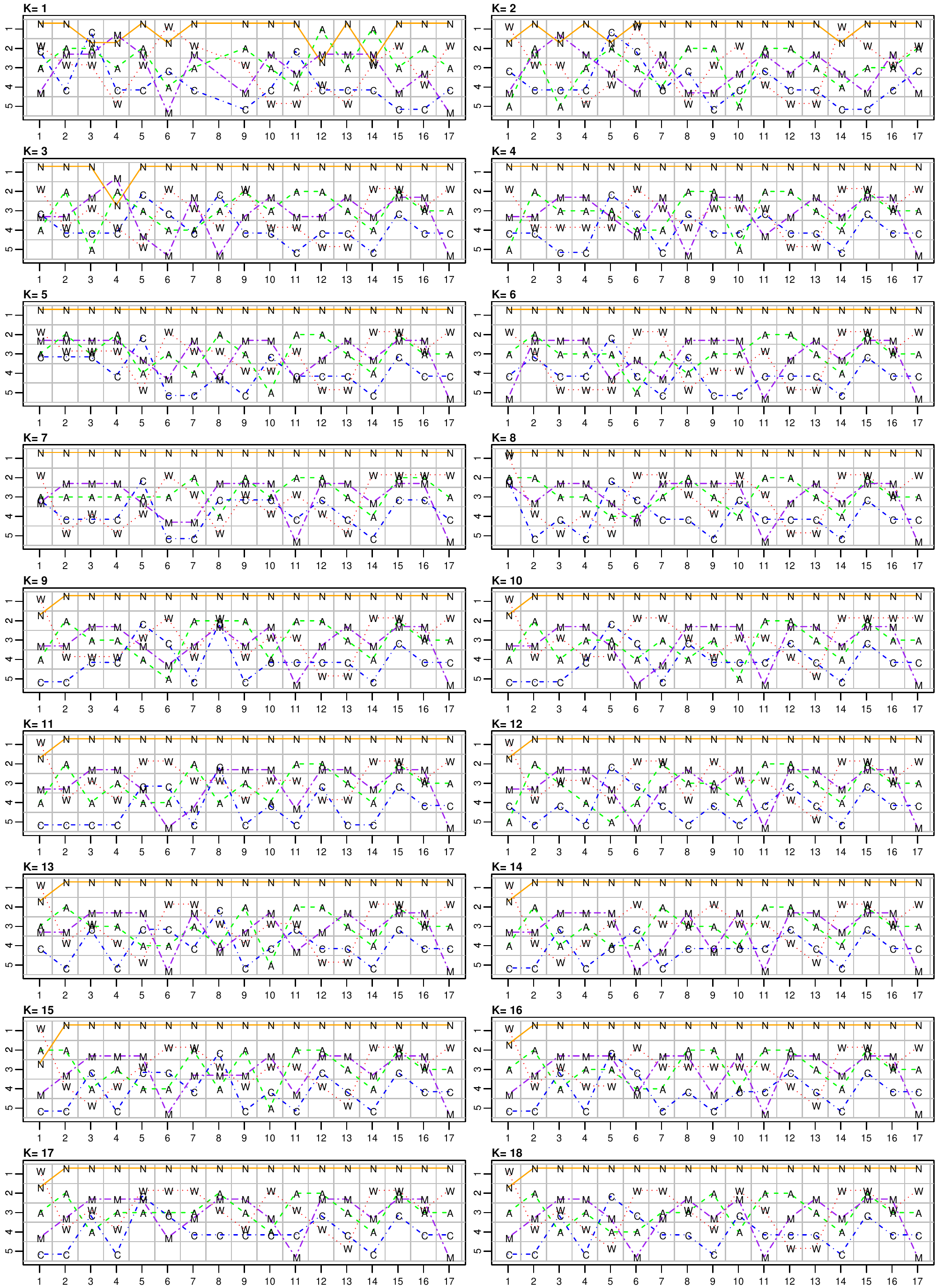}
		\caption{Variation in ranks of techniques evaluated using IC configuration. Each sub-figure represents a window size from (1 to 17), x-axis shows split point in time (1 to 17), y-axis shows the ranks of technique from (1 to 5), and K represents window size. Techniques: A=Amasaki15, W=Watanabe08, C=CamargoCruz09, N=Nam15, M=Ma12.}
		\label{fig:stepConfigIC}
	\end{figure*}

\begin{figure*}
		\centering
		\includegraphics[width=\textwidth]{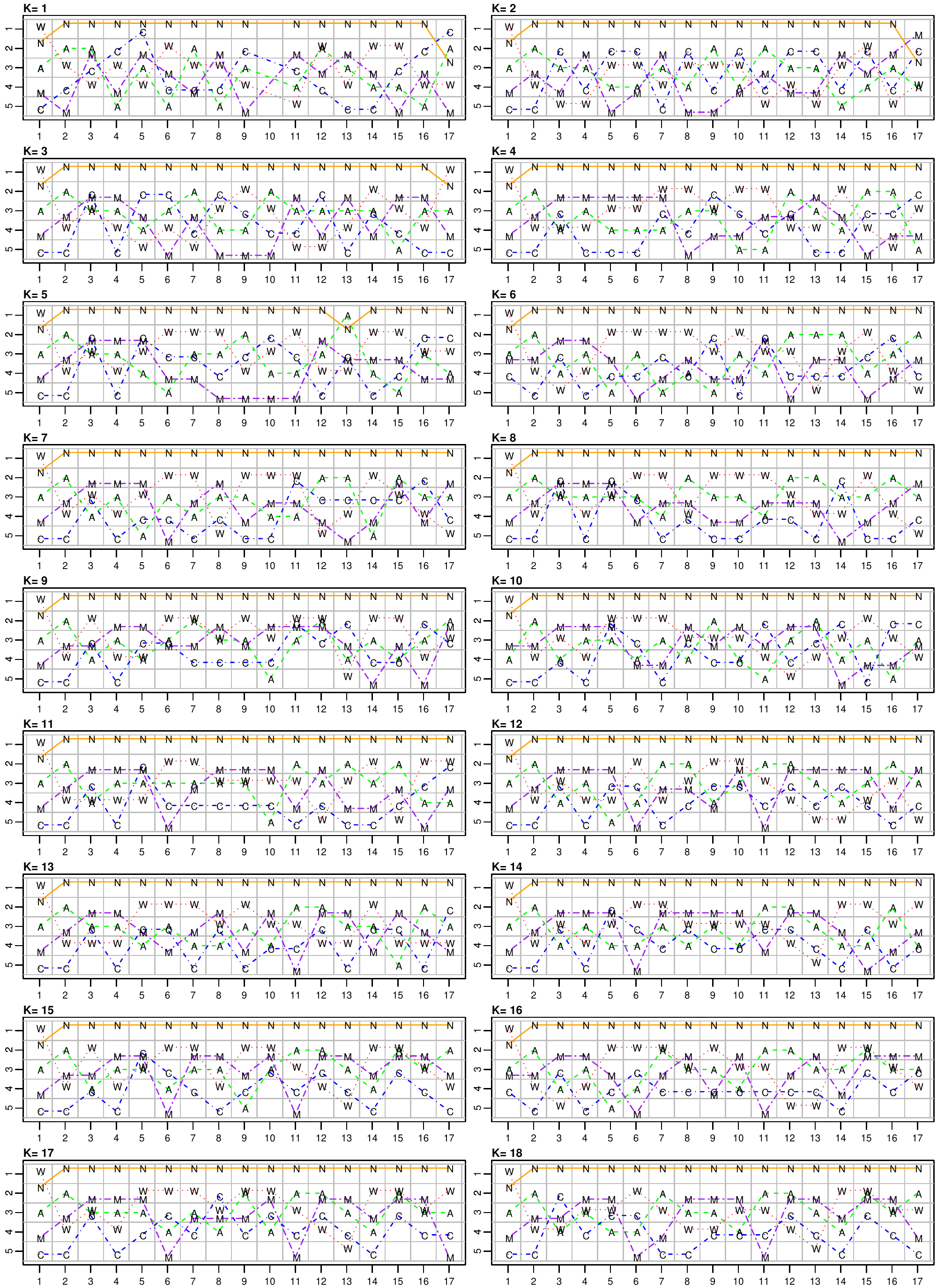}
		\caption{Variation in ranks of techniques evaluated using CI configuration. Each sub-figure represents a window size from (1 to 17), x-axis shows split point in time (1 to 17), y-axis shows the ranks of technique from (1 to 5), and K represents window size. Techniques: A=Amasaki15, W=Watanabe08, C=CamargoCruz09, N=Nam15, M=Ma12.}
		
		\label{fig:stepConfigCI}
	\end{figure*}

\begin{figure*}
		\centering
		\includegraphics[width=\textwidth]{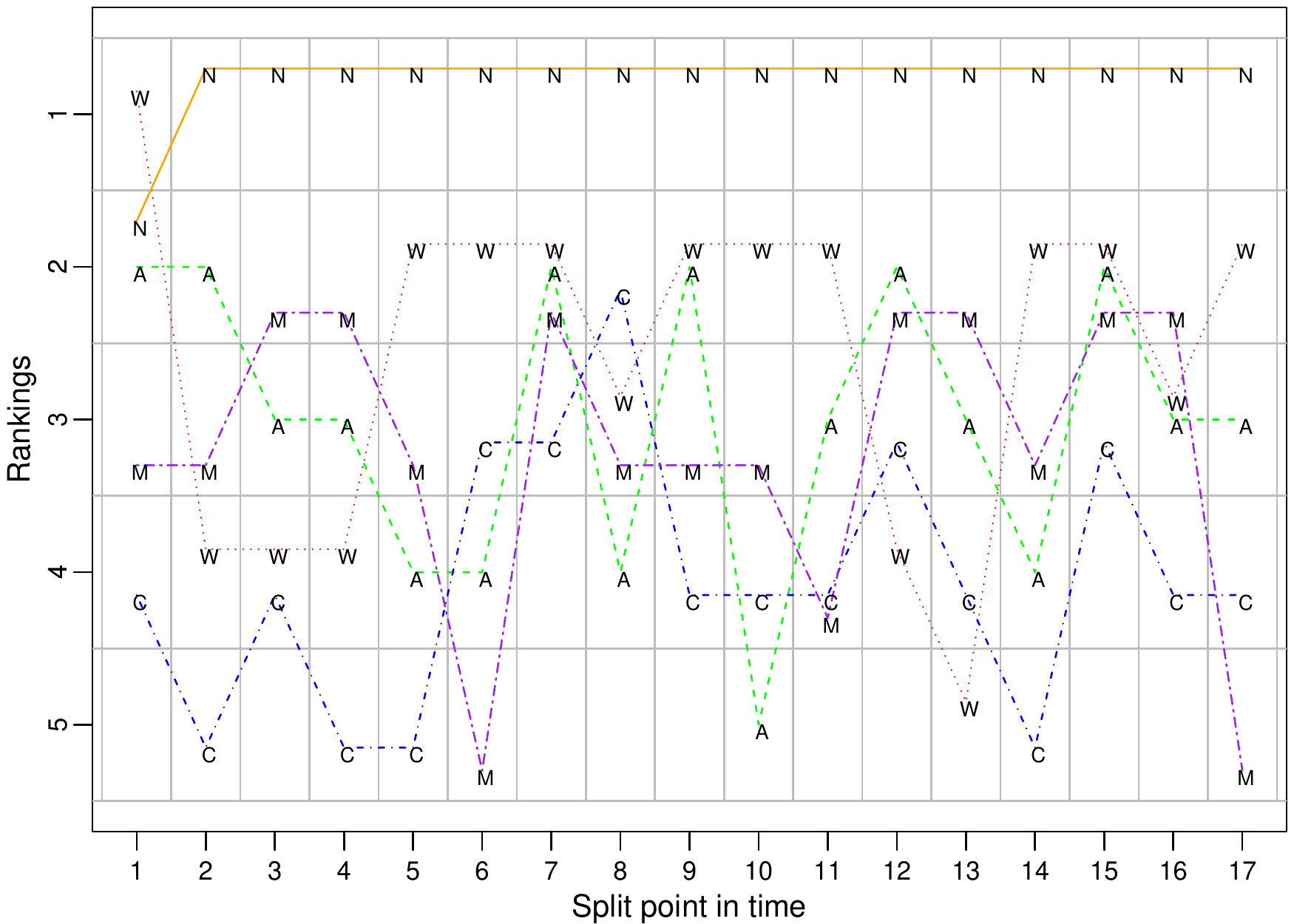}
		\caption{Variation in ranks of techniques evaluated using II configuration. The x-axis shows the split point in time (1 to 17), and the y-axis shows the ranks of technique from (1 to 5). Remember window size(K) does not matter in II. Techniques: A=Amasaki15, W=Watanabe08, C=CamargoCruz09, N=Nam15, M=Ma12.}
		\label{fig:stepConfigII}
	\end{figure*}

	The ranks of the evaluated tecniques vary in each configuration and four out of five techniques have a different rank in comparison with \textsc{HerboldMethod}. However, Nam15 which outperformed other approaches in \textsc{HerboldMethod} also obtained the top rank in all time-aware configurations. It is the only technique whose rank matches with \textsc{HerboldMethod} in addition to being consistent across the four configurations. Despite this one may observe an occasional decline for Nam15 at different split points in \figref{fig:stepConfigCC} through \figref{fig:stepConfigII}.
	Contrarily, the remaining four techniques, Amasaki15, Watanabe08, CamargoCruz09, and Ma12 remain inconclusive not just across configurations but also at different split points.
	
	\begin{table}[t]
		\begin{center}
			\caption{New ranks of techniques based on Mean Rank Score and their comparison with \textsc{HerboldMethod} ranks. For ease, the decimal values were replaced by whole numbers without affecting the ranks}
			
			\begin{tabular}{l>{\centering\arraybackslash}p{11em} cccc}
				
				\toprule
				
				\textbf{Technique}  &  \textbf{\textsc{HerboldMethod}} & \multicolumn{4}{c}{\bf New Ranks} \\
				& \textbf{Ranks}	& \textbf{CC} & \textbf{IC} & \textbf{CI} & \textbf{II} \\
				
				\cmidrule(lr){1-1} \cmidrule(lr){2-2}
				\cmidrule(lr){3-6} 
				
				Nam15 & 1 & 1 & 1 & 1 & 1 \\
				Ma12            & 2 & 2 & 3 & 3 & 4  \\
				Amasaki15   & 3 & 3 & 2 & 4 & 3 \\
				CarmagoCruz09       & 4 & 5 & 5 & 5 & 5 \\
				Watanabe08       & 5 & 4 & 4 & 2 & 2 \\
				
				\bottomrule
			\end{tabular}
			\label{table:newRanks}    
		\end{center}
	\end{table}
	\begin{table}
		\begin{center}
			\caption{Standard deviation in ranks of techniques calculated using Mean Rank Score of AUC, F-Score, G-measure and MCC metrics.}
			
			\begin{tabular}{lrrrr}
				\toprule
				\textbf{Technique} & \textbf{CC} & \textbf{IC} & \textbf{CI} & \textbf{II}  \\
				\midrule
				Ma12            & 1.04 & 1.07 &  1.07 &  0.99 \\
				Nam15           & 0.37 & 0.31 &  0.29 & 0.24 \\
				Amasaki15       & 0.99 & 0.88 &  0.90  & 0.93 \\
				Watanabe08       & 1.08 & 1.15 &  1.11   &  1.10\\
				CamargoCruz09   & 1.05 & 0.90 &  1.07  &  0.86 \\
				\bottomrule
			\end{tabular}
			\label{table:Ranking}
		\end{center}
	\end{table}
	
	To quantify the variation in the ranks of techniques, we present the standard deviation of ranks within each configuration in \tabref{table:Ranking}. The values of standard deviation range from 0.24 (smallest) in Nam15 to 1.15 (largest) in Watanabe08 which shows that the ranks of four techniques vary by at least $+/-1$ when evaluated at different time splits within a configuration. This variation shows that the performance of each technique varies depending on the context of evaluation and the ranks do not generalize over all time-periods.  
	
	\roundbox{
		According to the result of \textsc{HerboldMethod}, Nam15 outperforms the other techniques by achieving the first rank, whereas Watanabe08 performs. However, our evaluation shows that the ranks of all approaches not only vary at different split points within the configurations, but all except Nam15, have inconclusive ranks across the configurations as well.} 

	\section{Discussion}
	\subsection{Insights from Study}
	In this study, we show that defect prediction approaches can exhibit different performance when evaluated under different contexts. By using a subset of the Jureczko data used in the benchmarking study of \citet{Herbold}, we observed a disagreement with the ranks reported in Herbold's original study and \textsc{HerboldMethod}.
	
	We also explain in this paper that cross-validation is not an appropriate way of training cross-project defect prediction models because it randomly splits the data irrespective of time order. This type of evaluation might lead to the training of models on future data, which is in practice, not available for use at the time of prediction. As a result, the performance estimates of defect prediction models may be biased, and under realistic settings the model may perform better or worse than the estimates produced by making unrealistic assumptions.
	
	Studies in the past have engaged in time-travel because of a cross-validation based evaluation, therefore to avoid it, we adopt a time-aware evaluation, and report the standard deviation observed in the four performance metrics as well as the ranks of five techniques. A comparison of our resulting ranks with the ranks reported by \citet{Herbold} and the ranks obtained from \textsc{HerboldMethod} suggest that defect prediction models yield different conclusions when evaluated using time-aware evaluation and data from different time periods.
	
	In the context of time-aware evaluation, online defect prediction is the safest approach, because it trains the model on past data and evaluates it on future data. However, there is a difference between our proposed methodology and online defect prediction. Online defect prediction trains the model on complete data from the past, which is similar to our IC and II configurations. In contrast, in our CC and CI configurations, the model is trained on partial data from the past which is less resource intensive. 
	
	In the following sections, we consider some of the factors that may have caused the observed instability in the performance of defect prediction approaches.
	
	\subsection{Impact of factors other than time on conclusion stability}
	\label{sec:allfactors}
	
	\begin{figure*}
		\centering
		\includegraphics[width=\textwidth]{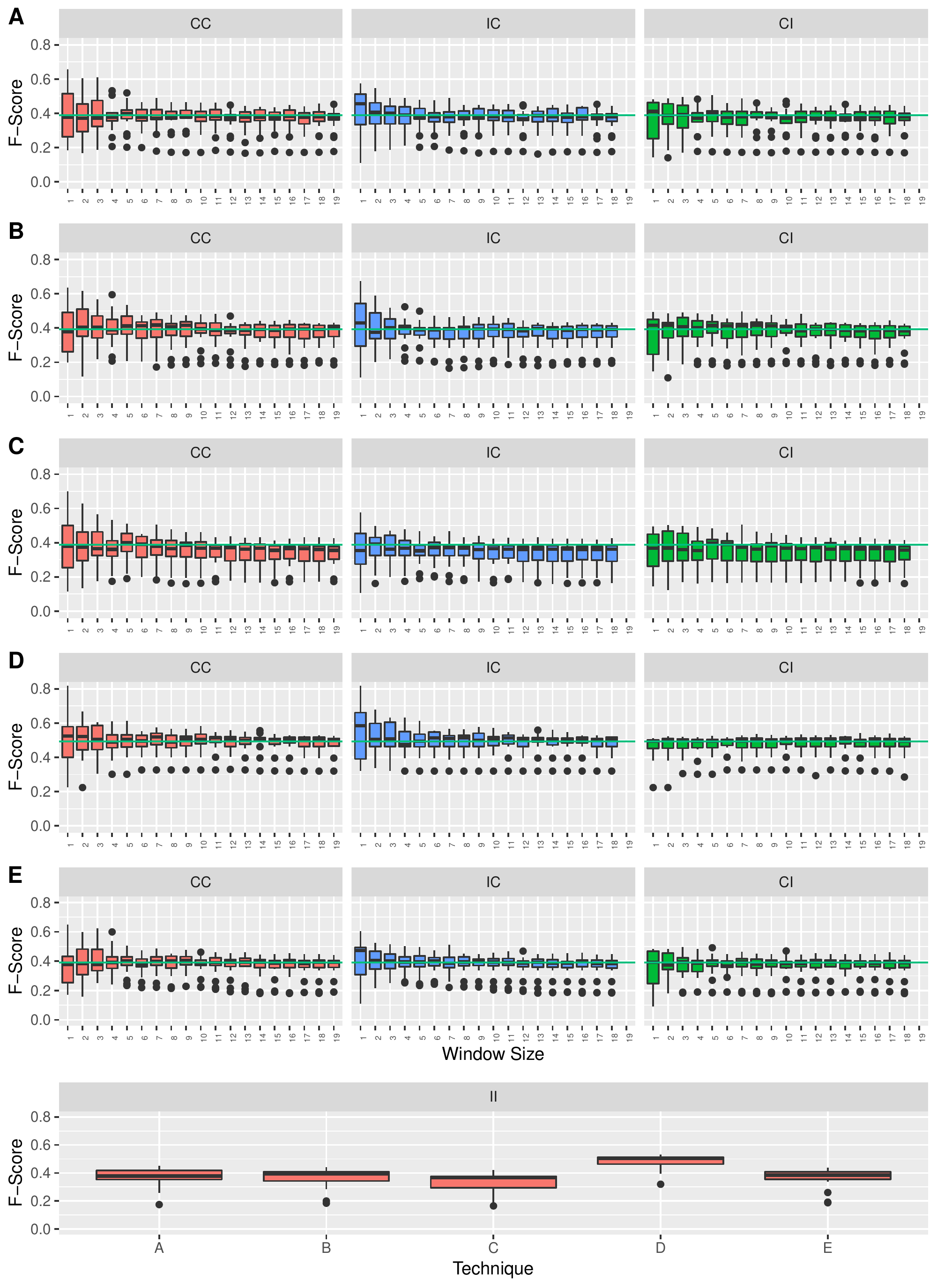}
		\caption{Comparison of F-Scores Standard Deviation (SD) of techniques A-Amasaki15, B-Watanabe08, C-CamargoCruz09, D-Nam15, E-Ma12. The Y-axis shows F-Score SD for a fixed Window Size (K). }
		\label{fig:windowFscoreSD}
	\end{figure*}
	\begin{figure*}
		\centering
		\includegraphics[width=\textwidth]{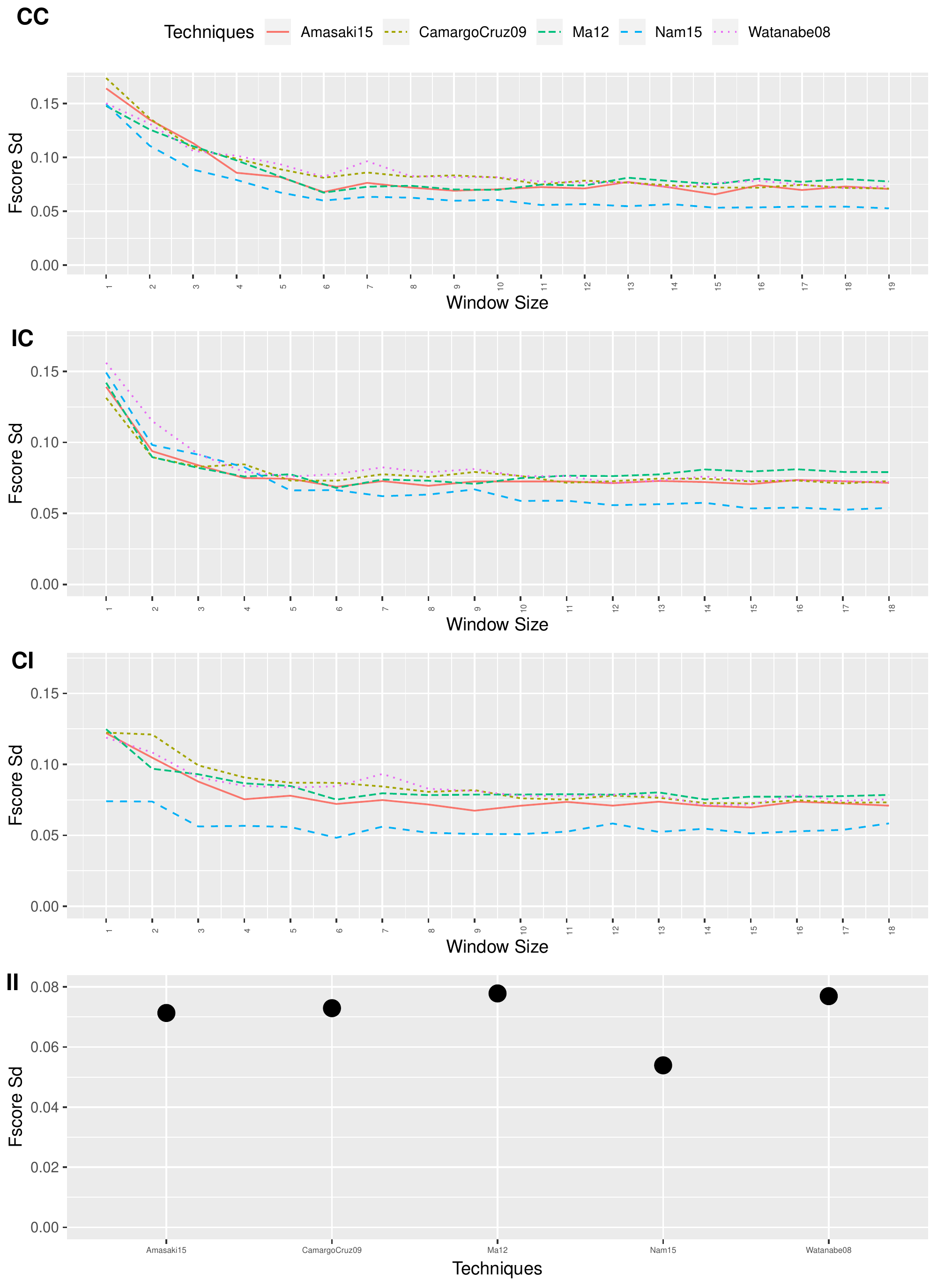}
		\caption{Standard deviation in F-Scores of five techniques at a given Window Size (K) show on X-axis. The Y-axis shows corresponding standard deviation in F-Score for that particular Window Size (K).}
		\label{fig:windowSD}
	\end{figure*}
	
	We observed that the standard deviation (SD) of F-Scores in \textsc{HerboldMethod} is high (i.e., $>0.1$). 
	Further, we compared the SD of \textsc{HerboldMethod}'s F-Score with the SD of time-aware configurations' F-Score. The technique wise F-Score SD of \textsc{HerboldMethod} is 0.132 for Amasaki15, 0.152 for Watanabe08, 0.159 for CamargoCruz09, 0.179 for Nam15, and 0.133 for Ma12. While in contrast to \textsc{HerboldMethod}, the technique wise F-Score SD of our time-aware configurations, as observed in \tabref{table:meanAndSdFscore}, is always $<0.1$. Hence, it is safe to assume that the newly reported ranks by the time-aware configurations in \tabref{table:newTechniqueValues} are more reliable than \textsc{HerboldMethod}.
	
	Having said that, there are several other possible factors that might have affected the conclusion stability of CPDP approaches. These factors include, but are not limited to, the following: noise in the dataset; types of the projects and stakeholders involved; software development process; the nature of the CPDP approaches themselves; uneven release distribution of projects over timeline; size of the dataset and imbalanced data classes etc. We investigate a few of the aforementioned factors below and their effects on the observed instability.
	
	\subsubsection{Impact of projects included in Tr-Test set:}
	\citet{Herbold} explored the impact of using a small subset
	of data on the model performance and found that it can lead to significantly
	different results. One might think that it is the case here as well because, in our data set, not all the projects are evenly spread across the timeline. For example, Xerces is only in the first few buckets (\figref{fig:timeBuckets}) and Camel is only in the last few buckets.
	As a result some instability may be caused due to the change of projects between different Tr-Test pairs. To counteract the effect of different projects on performance instability we divided the data into two subsets: Subset-1 includes buckets from 1990-07 to 2004-01, and Subset-2 includes buckets from 2004-07 to 2009-01. From the results presented in previous section, the CC configuration has shown highest variance, therefore we evaluate each subset by running only CC configuration. \tabref{table:subsetsComparison} shows that the F-Score standard deviation increased when we divided the \textsc{FilterJureczko} and both the subsets have a higher standard deviation when compared to the original evaluation on the entire data set. This confirms that the instability does not diminish even when same projects are evaluated over time. However,
	it is still hard to reason whether this difference is
	due to time-based evaluation or merely because the
	data set size has been further reduced.
	
	\begin{table}
		\begin{center}
			\caption{Comparison of F-Score Standard Deviation(SD) of \textsc{FilterJurezcko} with its two subsets. Subset-1 represents data from 1999-07 to 2004-01 and Subset-2 represents data from 2004-07 to 2009-01. The subsets are only evaluated for Configuration CC.}
			\begin{tabular}{lrrr}
				\toprule
				
				\textbf{Technique} & \textsc{FilterJurezcko} & \textbf{Subset-1} & \textbf{Subset-2}  \\
				\midrule
				Ma12            & 0.08 & 0.11  & 0.11 \\
				Nam15           & 0.07 & 0.09 &  0.12  \\
				Amasaki15       & 0.08 & 0.12 &  0.10   \\
				Watanabe08       & 0.09 & 0.12 &  1.11   \\
				CamargoCruz09       & 0.09 & 0.13 &  0.11   \\
				\bottomrule
			\end{tabular}
			\label{table:subsetsComparison}
		\end{center}
	\end{table}
	\begin{table}
		\begin{center}
			\caption{Standard Deviation(SD) of performance metrics when evaluated on balanced \textsc{FilterJurezcko} data set with CC configuration of time-aware methodology.}
			\begin{tabular}{lrrrr}
				\toprule
				
				\textbf{Technique} & \textbf{F-Score SD} & \textbf{MCC SD} & \textbf{AUC SD} & \textbf{G-measure SD} \\
				\midrule
				Ma12            & 0.13 & 0.24  & 0.12 & 0.12 \\
				Nam15           & 0.11 & 0.18 &  0.08 & 0.09  \\
				Amasaki15       & 0.12 & 0.19 &  0.10 & 0.09  \\
				Watanabe08       & 0.11 & 0.20 &  0.12 & 0.10  \\
				CamargoCruz09       & 0.11 & 0.21 &  0.11 & 0.10  \\
				\bottomrule
			\end{tabular}
			\label{table:sdOfBalancedCC}
		\end{center}
	\end{table}
	
	\subsubsection{Impact of data size:}
	\label{sec:constantWindowSize}
	The Tr-Test pairs generated using different window sizes (K) vary in terms of size i.e., the number of instances. The performance of a classifier can differ when trained using data sets of different scales. Consequently, the variation might seem to have been introduced due to the comparison between models trained using variable window sizes. To counteract this we fixed the window size while training prediction models and then compared the standard deviation for every window size individually. The variability in F-score across different values of K is shown in \figref{fig:windowFscoreSD}, and it can be seen that even for a fixed value of K, F-Score varies from 0.2 to 0.6 and occassionally 0.8. The standard deviation in F-score for a fixed value of K is also shown in \figref{fig:windowSD} and it can be concluded that despite a fixed value of K and data of similar scales, the instability is there and it remains high.
	
	\subsubsection{Impact of data imbalance:}
	Data imbalance refers to the unequal distribution of prediction class labels in the training data set. This imbalance may cause a defect prediction model to incorrectly classify between two classes during testing, which leads to inconsistent performance of the model across its different evaluations. To achieve a balanced distribution, prior work \citep{sampling2014,  zimmermann2007,kamei2016studying} has used several re-sampling techniques such over-sampling and under-sampling. Over-sampling uses the randomly selected minority class instances and adds them to the original data set. Under-sampling, on the other hand, removes random instances from the majority class until both classes become equal. 
	
	To balance our data set, we used the under-sampling methodology and then re-ran all five approaches using \textsc{HerboldMethod} and the CC configuration. We compared each technique's performance measures obtained using \textsc{HerboldMethod} with time-aware CC configuration using Wilcoxon rank-sum test. For all five evaluated approaches, the results of CC configuration still differ with \textsc{HerboldMethod} in a statistically significant way at an $\alpha=0.01$. \tabref{table:sdOfBalancedCC} shows that the standard deviation for all of the four metrics is above our threshold, showing that the instability in results cannot be clearly attributed to different class distributions.

	\subsection{Implications}
	Although, our study is limited to the area of cross-project defect prediction, the time-aware methodology employed in this paper can be used to evaluate the conclusion stability of other software analytic approaches, such as duplicate bug report prediction, effort estimation, and bad smell detection. To this end, our experimental results ascertain that our concern about over generalization of conclusions is legitimate. In our evaluation, which is based on four time-aware configurations, the ranks of techniques vary by +1 or -1 within the configurations as well as across them. Only Nam15 achieved the same rank in all four configurations and in the \textsc{HerboldMethod}. The other techniques degrade by 2 or 3 ranks in certain configurations, which means that there is no agreement and thus high instability in the remaining four ranks. On a side note, these configurations allow for a systematic way of generating training and test data and also seem promising, as evaluations based on them exhibit diverse results which are realistically closer to the performance that a technique will yield in practice.
	
	Lastly, it should be noted that the computational cost of training a large number of models corresponding to all configurations can be high, especially for models that employ sophisticated training techniques such as Neural Networks. Therefore, only some configurations or a few windows
	in each configuration may only be used to obtain realistic performance estimates. Having said that, the choice of configuration entirely depends on the purpose of evaluation, as we explained in the methodology section. In either case, however, a time-stamped data set or version release dates are required to carry out a more detailed evaluation, and therefore software engineering researchers who plan to collect defect prediction data in future shall also provide time information with their data sets.
	\section {Threats to Validity}

	\paragraph{Construct Validity.}
	We use the source code provided by \citet{herbold2017repKit} for the evaluation. This poses a threat to the construct validity of our study but to counteract that, we also look into the original papers and make sure the implementations were correct. 
	
	\paragraph{External Validity.}
	The external validity of the study is limited by the use of Jureczko data set. Our experiment relies on dates and timestamps which were not available in any of the publicly available data sets hence we relied on only a single data set for our study, Jureczko data set. The data set contains 20 metrics and the results of our study might only hold for data having similar characteristics. 
	
	Additionally, Jureczko data set does not contain bug-report and bug-fix timestamps. The data set was collected by analyzing the commit logs using a regular expression to decide if a commit is bug-fixing or not. Hence, we could not map release dates to bug report/fix times and as a result we might have time-traveled due to our ignorance of these. Although, it is out of the scope of our current study but in future we intend to update the bug-prediction data sets to associate bug information with commits.
	
	The standard deviation in the performance metrics of \textsc{HerboldMethod} and time-aware configurations does not necessarily suggest that it exists primarily due to time. Rather, there may be other factors affecting this standard deviation such as noise in the dataset, types of the projects, software development process, or the nature of the CPDP approaches themselves. 
	
	\paragraph{Internal Validity.}
	The internal validity of the study suffers to a small extent due to reliance on the assumptions made in prior works. We have not tuned the hyper-parameters of the decision tree but have instead relied on the evaluation settings similar to \citet{Herbold}. An interesting future work is to examine the effect of tuning model parameters on the results.
	
	
	\section{Conclusion}
	
	Software engineering researchers often make claims about the generalization of the performance of their techniques outside the contexts of evaluation.
	In this paper we investigate whether conclusions in the area of defect prediction---the claims of the researchers---are stable throughout time.  
	
	We show lack of conclusion stability for multiple techniques when they are evaluated at different points in a project's evolution. By following a time-aware methodology we found out that conclusions regarding ranking and performance of techniques replicated by \citet{Herbold} benchmarking study are not stable across different periods of time. With a standard deviation of $0.05$ or more in F-Score, MCC and G-measure, we find that with context (i.e., time) of evaluation, the relative performance of defect prediction techniques changes, provided the time frame and projects we used for evaluation. 
	
	However, it is hard to reason if time alone is the primary factor that leads to unstable conclusions, but our empirical evaluation shows that it does seem to be a factor. There may be other factors such as noise in the dataset, types of the projects, software development process, or the nature of the CPDP approaches themselves that require further investigation to determine their effect on conclusion stability.
	
	This case study provides evidence that in the field of defect prediction the context of evaluation (in our case, time) plays an important role.
	Therefore, it is imperative that empirical software engineering researchers do not over generalize their results but instead couch their claims of performance within the contexts of their evaluation---a field-wide faux pas that perhaps even this paper engages in.
	

	%
	%

	\bibliographystyle{spbasic}      
	\bibliography{References}
	
\end{document}